\documentclass[fleqn,usenatbib]{mnras}

\usepackage{newtxtext,newtxmath}
\usepackage[T1]{fontenc}

\usepackage{anyfontsize}

\DeclareRobustCommand{\VAN}[3]{#2}
\let\VANthebibliography\thebibliography
\def\thebibliography{\DeclareRobustCommand{\VAN}[3]{##3}\VANthebibliography}

\usepackage{graphicx}	
\usepackage{amsmath}	
\usepackage{pdflscape}
\usepackage{orcidlink}

\newcommand{\beq}{\begin{equation}}
\newcommand{\eeq}{\end{equation}}
\newcommand{\bea}{\begin{eqnarray}}
\newcommand{\eea}{\end{eqnarray}}
\newcommand\mstar{\mathrm{M}_{\bigstar}}
\newcommand\pcc{\rho_{xy}}

\newcommand{\bfx}{\mbox{\boldmath$x$}}
\newcommand{\bfk}{\mbox{\boldmath$k$}}

\newcommand{\bfq}{\mbox{\boldmath$q$}}

\title[Neural Network LRG Sampling]{Populating Large N-body Simulations with LRGs Using Neural Networks}

\author[M. Icaza-Lizaola et al.]{M. Icaza-Lizaola\,\orcidlink{0000-0002-2547-3184}$^{1,2,3}$\thanks{E-mail: miguel.deicaza@uam.es},
E.~L. Sirks\,\orcidlink{0000-0002-7542-0355}$^{4,5}$\thanks{E-mail: ellen.sirks@sydney.edu.au},
Yong-Seon Song\orcidlink{0000-0002-3261-6016}$^{1}$\thanks{E-mail: ysong@kasi.re.kr},
Peder Norberg\orcidlink{0000-0002-5875-0440}$^{6,7,8}$ 
\newauthor
\& Feng Shi\orcidlink{0000-0002-9968-2894}$^{9}$ \\
$^{1}$Korea Astronomy and Space Science Institute, 776 Daedeok-daero, Yuseong-gu, Daejeon 34055, Republic of Korea\\
$^{2}$Departamento de F\'isica Te\'orica,  Facultad de Ciencias M-8,   Universidad Aut\'onoma de Madrid, 28049 Madrid, Spain \label{inst1}\\
$^{3}$Centro de Investigaci\'{o}n Avanzada en F\'isica Fundamental (CIAFF), Facultad de Ciencias, Universidad Aut\'{o}noma de Madrid, 28049 Madrid, Spain\label{inst2}\\
$^{4}$School of Physics, The University of Sydney, Sydney, NSW, 2006, Australia\\
$^{5}$The Australian Research Council Centre of Excellence for Dark Matter Particle Physics, Australia\\
$^{6}$Institute for Computational Cosmology, Department of Physics, Durham University, South Road, Durham DH1 3LE, UK\\
$^{7}$Institute for Data Science, Department of Physics, Durham University, South Road, Durham DH1 3LE, UK\\
$^{8}$Centre for Extragalactic Astronomy, Department of Physics, Durham University, South Road, Durham DH1 3LE, UK\\
$^{9}$School of Aerospace Science and Technology, Xidian University, Xi'an 710126, China \\
}

\date{Accepted XXX. Received YYY; in original form ZZZ}

\pubyear{2024}

\begin{document}
\label{firstpage}
\pagerange{\pageref{firstpage}--\pageref{lastpage}}
\maketitle

\begin{abstract}

The analysis of state-of-the-art cosmological surveys like the Dark Energy Spectroscopic Instrument (DESI) survey requires high-resolution, large-volume simulations. However, the computational cost of hydrodynamical simulations at these scales is prohibitive. Instead, dark matter (DM)-only simulations are used, with galaxies populated a posteriori, typically via halo occupation distribution (HOD) models. While effective, HOD models are statistical in nature and lack full physical motivation.

In this work, we explore using neural networks (NNs) to learn the complex, physically motivated relationships between DM haloes and galaxy properties. Trained on small-volume, high-resolution hydrodynamical simulations, our NN predicts galaxy properties in a larger DM-only simulation and determines which galaxies should be classified as luminous red galaxies (LRGs).

Comparing the original LRG sample to the one generated by our NN, we find that, while the subhalo mass distributions are similar, our NN selects fewer low-mass subhaloes as LRG hosts, possibly due to the absence of baryonic feedback effects in DM-only simulations. This feedback could brighten or redden galaxies, altering their classification. 

Finally, we generate a new LRG sample by fitting an HOD model to the NN-generated LRG sample. We verify that both the HOD- and NN-generated samples preserve a set of bias parameter relations, which assume that the higher-order parameters, $b_{s2}$ and $b_{3\rm{nl}}$, are determined by the linear bias parameter $b_{1}$. These relations are commonly used to simplify clustering analyses.
\end{abstract}


\section{Introduction}

The study of the large-scale structure (LSS) of the Universe through galaxy surveys is one of the most significant endeavours in observational cosmology \citep[e.g.,][]{York_2000, Colless_2001, Eisenstein_2001, 2013AJ....145...10D, DESI_2016, Medinaceli_2022, 2022A&A...662A.112E}. These surveys map the positions of hundreds of thousands to millions of galaxies, allowing us to study their clustering and how it evolves over cosmic time. The observed clustering reflects the underlying distribution of matter, which is shaped by gravitational interactions and the relative densities of different energy components in the Universe. By comparing these observations with theoretical models, we can constrain key cosmological parameters and gain deeper insight into the processes driving cosmic structure formation.

The Dark Energy Spectroscopic Instrument \cite[DESI,][]{DESI_2016} is one of the largest LSS surveys to date and is currently in its fourth year of taking observations. DESI aims to observe different tracers of matter at different redshifts. In particular, DESI is building a catalogue of $\sim$eight million luminous red galaxies (LRGs) between  $0.4 < z < 1.0$ \citep{Zhou_2023}. LRG is the name given to very massive galaxies usually found at the centre of large galaxy clusters \citep{Kauffmann_2004}, making them strongly biased tracers of the underlying matter distribution. The centres of these cluster are environments with a high occurrence of mergers and interactions between galaxies which explains the massive sizes of LRGs. 

As their name suggests, LRGs have depleted most of their gas content, resulting in little to no ongoing star formation. Consequently, they appear redder than galaxies where young, blue stars are still being formed. The absence of younger stars produces a prominent spectral break at 4000\,Å, a sharp feature that allows LRGs to be efficiently identified in imaging surveys such as the DESI Legacy Imaging Surveys Data Release 9 \citep[DR9]{2019AJ....157..168D}. This feature also enables fast and reliable redshift estimations, even at high redshifts, making LRGs ideal tracers within LSS surveys. To select LRGs, DESI identifies targets using optical and near-infrared photometric data. A specific set of colour and magnitude cuts is applied to isolate LRG candidates within the desired redshift range.

The analysis of LSS surveys, such as the DESI LRG survey, depends on our ability to generate large and accurate simulated mock catalogues of galaxies that statistically reproduce their observed properties and spatial distribution. These mocks play a crucial role in testing data analysis methodologies, investigating systematic errors, and optimizing future survey designs.

Accurately modelling small-scale clustering requires high-resolution N-body simulations. However, when analysing clustering on larger scales, simulations must also cover correspondingly large volumes to minimize Poisson noise. Furthermore, to reduce the sample variance of clustering statistics, it is common practice to average over several realizations with different random initial conditions. The combination of large volumes, high resolution, and the need for multiple realizations leads to significant computational costs. For volumes exceeding a few hundred Mpc, incorporating hydrodynamical interactions becomes prohibitively expensive, making it impractical to simulate baryonic effects at these scales. Consequently, simulations with the volume and resolution necessary for LSS surveys like DESI only model the dark matter (DM) distribution \citep{2021MNRAS.508.4017M}, requiring galaxies to be added to the DM overdensities in post-processing.

One of the standard approaches to populate simulations with galaxies is  through the halo occupation distribution (HOD) methodology \citep[e.g.][]{2000MNRAS.318.1144P,Zheng_2005,2021MNRAS.502.3582Y} which makes the assumption that all galaxies live inside a DM halo, and provides a statistical description for the probability of finding one or more galaxies within a specific DM halo. The probabilities are usually based on a set of halo properties. More traditional approaches consider only the halo mass \citep[e.g.][]{Zheng_2005}, however, newer and more complex models consider other properties correlated with the assembly history of the halo, the {\it assembly bias} \citep[e.g.][]{2024MNRAS.530..947Y}. HOD models have been used extensively in preparation to the first data release of DESI, in particular, to populate the AbacusSummit N-body suite with DESI LRG mocks \citep{2024MNRAS.530..947Y}, and to test the DESI LRG pipelines \citep{2024arXiv241112025B}. 

HOD models rely on a set of adjustable free parameters that describes the halo occupation statistic. To determine the HOD parameters one typically uses the values that best reproduce a set of observed survey statistics, commonly the small scale projected two-point correlation function. However, it is important to note that these methods are statistical and do not attempt to directly model the physical processes shaping the connection between galaxies and DM haloes. While optimizing these parameters one ensures that mock galaxy samples exhibit the correct statistic that it was fitted to when training, with the statistic representing a condensed version of the data. Consequently, when using HODs, there is an inevitable loss of information.

In this work, we propose an alternative approach to populating DM haloes with galaxies using a physically informed method based on neural networks (NN), a type of machine learning (ML) algorithm. For each DM subhalo, we aim to accurately predict the mass, colours, and magnitudes of the galaxy it hosts, allowing us to determine whether it meets the LRG selection criteria. These NN models are trained on a small hydrodynamical simulation. Subsequently, we apply these NN models to populate LRGs in a large DM-only N-body simulation.

There have been several works that attempt to use ML to predict galaxy properties from their host halo or subhalo properties inside a hydrodynamical simulation \citep[e.g.][]{Kamdar_2016,Agarwal_2018,de_Santi_2022,Lovell_2022,Chuang_2023,Icaza-Lizaola_2023,Dai_2024}. It is generally acknowledged that modelling the stellar mass can be done accurately from subhalo properties alone \citep[e.g.,][]{Kasmanoff_2020, Icaza-Lizaola_2021, Icaza-Lizaola_2023}. This can be attributed to the strong correlation between the mass of a host subhalo and the stellar mass of a galaxy \citep{1978MNRAS.183..341W}. The remaining scatter in the stellar mass-halo mass relation is found to be well-correlated with the assembly history of the subhalo \citep{10.1111/j.1365-2966.2004.07733.x,10.1111/j.1745-3933.2005.00084.x,10.1111/j.1745-3933.2007.00292.x,2019MNRAS.489.2977R}. Therefore, models that utilize parameters correlated with the subhalo mass and the growth history of the subhalo can predict stellar mass with great accuracy.

However, attempts to model other properties like colours or magnitudes of galaxies have been much less successful, leading to larger discrepancies between the predicted and true colours of galaxies in test sets \citep[e.g.,][]{Kamdar_2016, de_Santi_2022, Chuang_2023}. This lack of accuracy is, at least partially, attributed to colours and magnitudes being influenced by different baryonic processes absent in DM-only simulations. The colours of galaxies serve as a proxy for their stellar composition, which, in turn, is dependent on phenomena such as the star formation rate, supernova feedback, AGN feedback and many other baryonic processes. Importantly, these phenomena are not necessarily well-correlated with the DM properties of host haloes, making the modelling of colours in general much more challenging. 

As a consequence, a NN model that selects LRG candidates using colour and magnitude cuts will include some contamination from galaxies slightly less red or bright than the imposed cut. In this work we explore how accurately our NN can reproduce the original LRG sample despite these difficulties.

Unlike NN models, which attempt to learn complex, and ideally physically motivated, relationships from training data, HOD models rely on statistical prescriptions to populate haloes with galaxies. Typically, it is challenging to assess how well HOD-generated galaxy samples recover the galaxy-halo connection of the original dataset they were fitted to. This difficulty arises because observational surveys cannot directly observe the host haloes of galaxies. As a result, when applying HOD analysis, one normally lacks an independent dataset for validation.

However, our NN LRG sample puts us in a unique position to address this issue: since it contains both haloes and galaxies, we can apply an HOD analysis to this sample and directly compare the galaxy-halo connection in the resulting HOD galaxy sample with that of the original NN LRG galaxy sample.

In particular, we are interested in understanding how HOD modelling influences the theoretical description of the galaxy-halo connection. In clustering analyses, it is common practice to construct a galaxy sample and compare its clustering properties with theoretical predictions. A standard approach is to incorporate the galaxy-halo connection into theoretical modelling through, e.g.\, second-order perturbation theory \citep{2015MNRAS.451..539G,2019JCAP...06..013Z}. This introduces additional nuisance parameters that must be fitted alongside the cosmological parameters. These parameters are commonly referred to as {\it galaxy bias} parameters. In clustering analyses, it is generally assumed that only a subset of these bias parameters are independent, while the rest are fully determined by the independent ones \citep{2009JCAP...08..020M}.
Within the DESI collaboration, parameter compression is routinely applied in redshift-space distortion (RSD) analyses \citep[e.g.,][]{2024PhRvD.110f3538I, 2025JCAP...01..129R} to reduce the number of free parameters, thereby minimizing prior effects and simplifying the analysis. These consistency relations between bias parameters are predicted by perturbation theory and validated using galaxy mock maps generated through HOD models \citep[e.g.][]{2024PhRvD.110f3538I, 2024arXiv240912937Z}. In these relations most higher-order galaxy bias parameters are expressed in terms of the leading-order bias parameter. Here, we investigate whether these consistency relations hold when using alternative galaxy mock catalogues generated trough physically informed ML methods. This is particularly important, as prior assumptions about higher-order galaxy bias parameters, usually considered nuisance parameters, could introduce biases in the actual cosmological parameters of interest.

In this work, we apply an HOD methodology to populate the MDPL2 DM haloes and fit the HOD parameters to reproduce the small-scale clustering of our NN LRG sample. We then independently test the relations between bias parameters in both models and examine whether the dependencies observed in HOD modelling also hold in our original NN LRG sample, which is designed to be more physically motivated.

This paper is organised as follows. Section~\ref{sec:data_TNG} presents the simulation data used for training our NN and describes how we select an LRG sample based on galaxy properties. We also introduce the set of subhalo properties used to train our NNs and the galaxy properties we aim to model. In Section~\ref{sec:NN_train}, we describe our NN-based framework for populating LRGs, including the prediction of galaxy colours and magnitudes, LRG selection, and the evaluation of how well the generated sample matches the original.

In Section~\ref{sec:data_MDPL2}, we introduce our large-volume DM-only simulation, and describe the galaxy sample obtained when applying our NN model to populate this simulation. Section~\ref{section4} then explores the use of this NN-generated LRG sample to test HOD models and assess their ability to recover the galaxy-halo connection. We begin in Section~\ref{HOD_formalism} by introducing our HOD formalism. Sections~\ref{Fitting_HODs}, \ref{HOD-NN}, and \ref{HOD_bestfit} then outline the methodology for fitting the HOD parameters to reproduce the statistical properties of our NN LRG sample and compare the resulting LRG population to the original sample. In Section~\ref{PT_work}, we introduce a theoretical galaxy-halo model, examine the relationships between its free parameters, and assess whether these parameter dependencies remain consistent between the HOD-generated sample and the original NN sample. Finally, in Section~\ref{conclussions}, we summarize our conclusions.

\section{Learning how to find LRG host haloes}
\label{sec:data}

In this section, we describe the training data used for our NN and the key properties of interest. We discuss how LRGs are defined within this dataset and then introduce our NN model. Finally, we explain the method used to select LRGs from the NN’s predictions.

Once completed, the DESI LRG catalogue will contain data on $\sim$eight million LRGs between $0.4 < z < 1.0$ \citep{Zhou_2023}. The median redshift of this sample is approximately $z\approx0.82$. As such, for all the simulation data used in this project, we use the snapshots that are as close as possible to this redshift. When considering our training properties, we only consider the snapshots up to $z=0.82$.

\subsection{Training data: the IllustrisTNG project}\label{sec:data_TNG}
For the training data of our NN, we use the IllustrisTNG suite of cosmological simulations \citep{Nelson_2019}. These are large volume, cosmological, gravo-magnetohydrodynamical simulations run with the moving-mesh code AREPO \citep{Springel_2010}. The TNG project is the successor of the original Illustris simulation and its associated galaxy formation model. TNG incorporates an updated galaxy formation model including new physics and numerical improvements, as well as refinements to the original model. 

The three flagship runs of IllustrisTNG are each accompanied by lower-resolution and DM-only counterparts. Each hydrodynamical TNG simulation solves for the coupled evolution of DM, cosmic gas, luminous stars, and supermassive black holes from a starting redshift of $z = 127$ to the present day, $z = 0$. 

We exploit the biggest simulation box of the TNG project with a side length of 205\,Mpc\,$h^{-1}$, roughly 300\,Mpc, referred to hereafter as TNG300. We consider the highest resolution versions of the hydrodynamical and the DM-only runs, which correspond to a DM particle mass of $5.9\times10^{7}$ and $7.0\times10^{7}\,\mathrm{M}_{\odot}$ respectively. The initial gas particle mass of the hydro simulations is $1.1\times10^{7}\,\mathrm{M}_{\odot}$. Both simulations assume the cosmological parameters of \citet{Planck_2016}: $\Omega_\mathrm{m}=0.3089$, $\Omega_\mathrm{b}=0.0486$, $h=0.6774$, $n_\mathrm{s}=0.9667$ and $\sigma_{8}=0.8159$. We use snapshot 55 of the DM-only and hydrodynamical run which has a redshift of $z=0.82$ in both simulations.

Groups of particles are detected using a standard \textsc{Friends-of-Friends} \citep[FoF, ][]{Davis_1985} algorithm with a linking length of $b = 0.2$. The FoF algorithm is run on the DM particles, and the other types (gas, stars, black holes) are attached to the same groups as their nearest DM particle. Individual subhaloes are identified using the \textsc{subfind} \citep{Springel_2001, Dolag_2009} algorithm. When identifying gravitationally bound substructures, \textsc{subfind} considers all particle types. To avoid ambiguity, we will refer to parent FoF groups as haloes and DM haloes of individual objects as subhaloes. 

Throughout this work, the total mass, M$_{\mathrm{tot}}$, of every subhalo (both centrals and satellites) is defined as the total mass of all particles gravitationally bound to it as calculated by \textsc{Subfind} (which would only include DM particles in the TNG300 DM-only box). In the case of the hydrodynamical simulation, we define stellar mass, $\mstar$, as the total mass of all stellar particles bound to the subhalo.

Merger trees have been created for the TNG simulations using \textsc{SubLink}\footnote{LHaloTree merger trees also exist. These two methods should on average give the same result, however, we choose to use the merger trees generated with \textsc{SubLink}.}, described in the \textit{Sussing Merger Trees} comparison project \citep{Srisawat_2013, Avila_2014, Lee_2014}. A subhalo in a given snapshot can have multiple progenitors in the previous snapshot, but we define the main progenitor as the one for which the mass summed across all earlier snapshots is the largest. We use the main branches of subhaloes to trace their properties through time, where the main branch of a subhalo is comprised of its main progenitors and descendants. 

\subsubsection{Matching subhaloes between simulations}\label{sec:matching}

In our work, we use a set of models trained on the TNG300 hydrodynamical simulation to populate the MDPL2 DM-only simulation. However, this exercise requires caution as the DM haloes in the hydrodynamical simulation might be influenced by baryonic processes, resulting in slight differences from their counterparts in the DM-only simulations. It is well known that baryonic processes can alter various properties of DM haloes, such as their shape \citep{1991ApJ...377..365K,2013MNRAS.429.3316B} or density profile \citep{2012MNRAS.422.3081M,2015MNRAS.452..343S}.

A common approach to address these issues is by matching the DM (sub)haloes between a hydrodynamical simulation and a DM-only simulation run using the same code and covering the same volume. The simulation settings should be identical except for the presence of baryons. In this work, we match subhaloes between the hydrodynamical and DM-only runs of TNG300. Since these two simulation boxes have the same initial conditions, they produce nearly the same DM (sub)halo populations. A \textit{Subhalo Matching To Dark} catalogue is readily available for TNG300. This catalogue is generated using the procedure of \citet{Nelson_2015}. The matching is performed based on the \textsc{LHaloTree} matching algorithm, and is bidirectional. By comparing DM particle IDs, which are unique, a matched subhalo in the other simulation is defined as the one with the highest fraction of common particles. The match must be bidirectional, i.e.\ if a subhalo in the hydrodynamical simulation is matched to a given subhalo in the DM-only box, then in turn the DM-only subhalo must have the original hydrodynamical subhalo as its best match. When necessary, FoF haloes are matched based on their central subhalo.

\subsubsection{Baryonic properties}
\label{sec:prediction_params}

In what follows, we introduce three baryonic properties. Our aim is to build a NN that can accurately model these properties, as they are later used to select LRG hosts within our methodology. These properties are the stellar mass, $z$-band magnitude, and $(r-z)$ colour.

In TNG300, the stellar mass of a galaxy, $\mstar$, is defined as the sum of all stellar particles which are gravitationally bound to the given subhalo as determined by \textsc{Subfind}. We will use this definition for stellar mass throughout. 

We use mock galaxy $r$- and $z$-band magnitudes calculated by \cite{2022MNRAS.512.5793Y} for the TNG300 galaxies. In summary, the stellar particles of each subhalo are divided into a young and old sample, based on the reported value of their star formation history. Then the Flexible Stellar Population Synthesis code \citep[\textsc{FSPS};][]{2010ascl.soft10043C,2010ApJ...712..833Cc} uses this together with the metallicity information of the particles to compute the stellar continuum of each population. \cite{2022MNRAS.512.5793Y} also include an empirical model of the optical depth that corrects the magnitudes for dust attenuation. A more detailed description of the galaxy magnitude models can be found in \cite{2021MNRAS.502.3599H}.

Most likely there are some systematic differences between how magnitudes are estimated for the DESI target selection of \cite{2020RNAAS...4..181Z} and how they were modelled by \cite{2022MNRAS.512.5793Y} for TNG300. To account for such differences, \cite{2022MNRAS.512.5793Y} allow for a small correction to their nominal TNG300 galaxy magnitudes: $z^{\prime}=z+\delta_{m}$ and $r^{\prime}=r+\delta_{m}$. To match the number density of selected LRGs to the expected density reported in \cite{2020RNAAS...4..181Z}, they fix $\delta_{m}=-0.4$. We use these corrected magnitudes in this work and will from here on refer to them as $r^{\prime}$ and $z^{\prime}$. For clarity, we refer to the colour as $(r^{\prime} - z^{\prime})$, but note that since the magnitude shift is band independent, the galaxy colour remains unaffected by this $\delta_{m}$ correction.

\cite{2022MNRAS.512.5793Y} generated colours for 80,000\footnote{There are actually 82,590 galaxies in this stellar mass range. We comment on the missing subhaloes in Appendix~\ref{sec:appendixA}.} galaxies in TNG300, with a stellar mass between $10 < \log_{10}(\mathrm{M_{\bigstar}}/(\mathrm{M_{\odot}}h^{-1})) < 12$. The higher mass cut discards the 42 most massive galaxies of TNG300 and was initially introduced in the sample as it was designed for studying emission line galaxies, that tend to be inside smaller haloes than LRGs. We describe how we deal with these missing galaxies within our methodology in Section~\ref{sec:LRG_selection}.

\subsubsection{DM-only properties}
\label{sec:training_params}

In this section we describe the DM-only parameters that are used by our NN to predict the baryonic properties described in Section~\ref{sec:prediction_params}. We do not make use of all possible subhalo parameters available in TNG300, but only those parameters that are also available in the MDPL2 DM-only simulation that we wish to populate. We note that there are additional properties that could be included and that might improve the accuracy of our NN predictions, but we defer this to a future study.

We make use of \textit{nine} parameters: whether the subhalo is a central or a satellite subhalo; the maximum velocity of the rotation curve V$_\mathrm{max}$ at $z = 0.82$; the total subhalo mass M$_\mathrm{tot}$ at $z = 0.82$, M$_{z=0.82}$; the maximum mass the subhalo has reached at any point in its past evolution, M$_\mathrm{max}$; the redshift at which the M$_\mathrm{max}$ is reached; the four redshifts at which the subhalo reaches four specific fractions of its maximum subhalo mass.

\begin{figure}
\centering
\includegraphics[width=\linewidth]{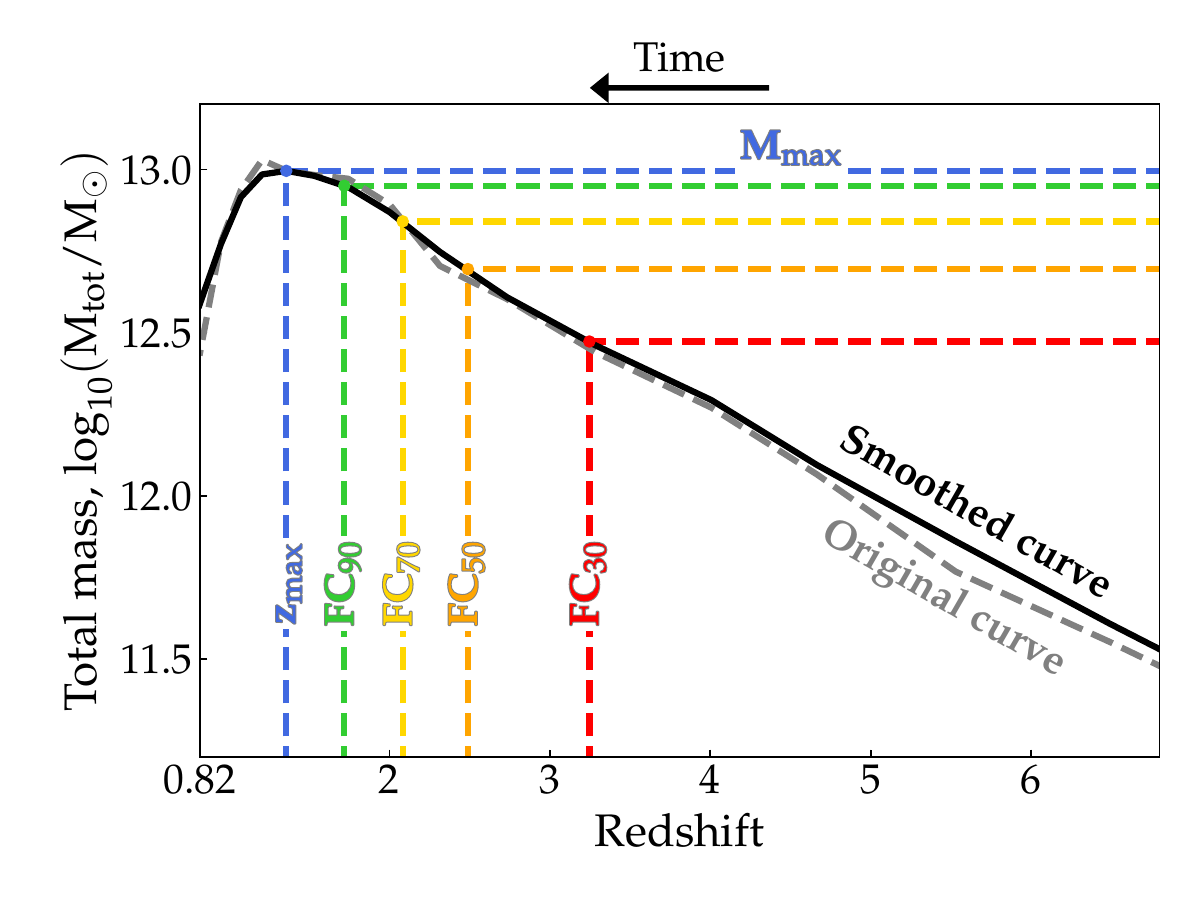}
\caption{Example of how the formation criteria described in Section~\ref{sec:training_params} are determined. The figure shows total mass as a function of redshift. First this mass curve is smoothed using a Gaussian kernel. Then the maximum mass, M$_\mathrm{max}$, is determined. The formation criterion is defined as the redshift at which the subhalo mass has reached a given fraction of M$_\mathrm{max}$, i.e.\ FC$_{70}$ is the redshift at which the subhalo has reached 70~per~cent of its M$_\mathrm{max}$. We only include redshifts satisfying $z \geq 0.82$. As redshift increases from left to right, time increases from right to left.}
\label{fig:formation_criteria}
\end{figure}

To find the redshift at which a subhalo reaches its maximum mass M$_\mathrm{max}$ (and fractions of this maximum mass), we consider the evolution of the total mass up until redshift $z=0.82$. We first smooth this mass versus redshift curve using a Gaussian kernel with a smoothing length of one standard deviation, making use of the Python function \textsc{astropy.convolution.Gaussian1DKernel}. We then identify the redshift at which a given subhalo has formed its maximum mass, and in turn certain fractions of its maximum mass. In particular, we consider the fractions $f = [0.3, 0.5, 0.7, 0.9]$. We refer to the corresponding redshifts from here on out as formation criteria FC$_{[30, 50, 70, 90]}$  (see \citealt{Icaza-Lizaola_2021,Icaza-Lizaola_2023}). Fig.~\ref{fig:formation_criteria} illustrates this method for one of our most massive subhaloes. We track the evolution of the subhaloes down to $z=10.98$ (snapshot 3), ensuring that we can identify when 99.8~per~cent of the subhaloes have formed 30~per~cent of their maximum mass, such that that FC$_{30}$ is well defined. We find that on very rare occasions the ordering of FC$_{30}$ $>$ FC$_{50}$ $>$ FC$_{70}$ $>$ FC$_{90}$ is not preserved due to e.g.\ merger events or possibly misidentification from \textsc{Subfind}. However, this only affects approximately 0.2~per~cent of the subhaloes, and so we leave these in our sample.

The nine properties introduced above are selected because they provide crucial information about the main characteristics and evolution of the subhalo up to its current stage and should be correlated with the galaxy properties we are modelling. The total subhalo mass M$_\mathrm{tot}$ serves as a reliable indicator of the galaxy size through the stellar-halo mass relation. V$_\mathrm{max}$ is correlated with the halo size but is less influenced by interactions in the halo's outskirts, making it a valuable tracer of the historical size of galaxies. \cite{Lovell_2022} have shown V$_\mathrm{max}$ to be more effective than M$_\mathrm{tot}$ in modelling certain galaxy properties when evolutionary information is absent, although in \cite{Icaza-Lizaola_2023} we find that the additional information it contributes to models of stellar mass its minor when the formation criteria parameters are included.

The assembly history of the halo can explain part of the scatter in the stellar-halo mass relation as galaxies that have been massive for a longer period have had an historically larger strong gravitational potential well, allowing them to grow larger than subhaloes of the same mass that only recently grew to their current size. To incorporate this information into our model, we track the redshifts at which our haloes grew into different fractions of their maximum size through the four formation criteria parameters.

Finally, we note that the evolution of central and satellite subhaloes differs significantly. Central subhaloes generally experience continuous growth throughout their evolution, while satellite subhaloes tend to decrease in size due to tidal interactions after becoming subhaloes of larger systems. Our model incorporates this information by including both a central/satellite flag and the M$_\mathrm{max}$ parameter. For central galaxies, M$_\mathrm{max}$ is likely very similar to M$_\mathrm{tot}$. However, for satellite galaxies, M$_\mathrm{max}$ provides additional information. Subhaloes that were significantly larger in the past had a deeper gravitational potential well, which facilitated the formation of massive galaxies. This distinction allows our model to account for the different evolutionary paths and physical characteristics of central and satellite subhaloes. We also include the redshift at which the maximum subhalo mass is reached as another free parameter to help differentiate the evolution between centrals and satellites.

We highlight that some of these parameters are strongly correlated with each other and may contain overlapping information. For instance, both M$_\mathrm{max}$ and V$_\mathrm{max}$ are indicators of the subhalo's size and can reflect similar underlying properties of the galaxy. This redundancy can be problematic in some statistical models, where multicollinearity can lead to less reliable inferences. However, this is not an issue for NN models, which can learn to identify any redundant information, and extract the most relevant features.

As stated in Section~\ref{sec:data_TNG}, the total mass of subhaloes in the TNG300 DM-only box consists of all DM particles gravitationally bound to the subhalo. The DM-only simulation box MDPL2, that we wish to populate with LRGs, uses a slightly different total subhalo mass definition, as discussed further in Section~\ref{sec:MDPL2_sample}.

\subsubsection{Selecting LRGs in TNG300}
\label{sec:LRG_selection}

In order to select observational targets for the LRG sample, DESI uses photometric data from various optical and infrared bands, as described by \cite{2019AJ....157..168D}. Magnitudes and colours derived from these bands are then used to determine targets based on several selection criteria \citep{2020RNAAS...4..181Z,Zhou_2023}. Among these criteria, the one relevant for our LRG selection within the TNG300 sample is the application of colour and luminosity cuts, which identify bright red galaxies as potential LRG candidates.

In the context of DESI, and throughout this work, we use the $(r^{\prime}-z^{\prime})$ colour and $z^{\prime}$-band magnitude cuts of \citet{2020RNAAS...4..181Z}. In the redshift range of interest for this study (i.e.\ $z\sim 0.8$), LRGs are selected through:

\begin{equation}
\label{eq:colour_mag_cuts}
(r^{\prime} - z^{\prime}) > (z^{\prime} - 16.83) \times 0.45
\end{equation}

Since massive galaxies tend to be bright in their rest-frame, they should be bright in the $z$-band. On the other hand, the extent of a galaxy's redness can be quantified through its $(r-z)$ colour. Therefore in order to select TNG300 LRGs we utilize the $r^{\prime}$ and $z^{\prime}$ introduced in Section~\ref{sec:prediction_params}.

From now on, we consider all TNG300 galaxies for which their colour satisfies the condition of equation~\eqref{eq:colour_mag_cuts} as \textit{true} LRGs. Our main goal is to use NNs to try to predict this colour and apparent magnitude and use the predictions to determine which subhaloes should contain a \textit{true} LRG. This can be done either by modelling $(r^{\prime}-z^{\prime})$ and $z^{\prime}$ directly, or by modelling correlated variables that are easier to determine trough DM-only properties like $\mstar$, as discussed in Section~\ref{sec:NN_train} below.

We stated that our mock predictions from the $r$- and $z$-band magnitudes were generated for galaxies within the $12 > \log_{10}(\mathrm{M_{\bigstar}}/(\mathrm{M_{\odot}}h^{-1})) > 10$ stellar mass range. As a consequence, we have no colour estimates for the 42 most massive galaxies in our sample. All of these 42 galaxies are inside very massive host subhaloes, with the smallest of them having a subhalo mass of $\log_{10}(\mathrm{M_{tot}/M_{\odot}})=13.8$. Additionally, 41 of the 42 galaxies live in the central subhalo of a very large halo. In general, we expect large galaxies in the centre of massive haloes to be LRGs. With this in mind, and throughout this work, we consider these 42 galaxies as part of our \textit{true} LRG sample, and they will be used when comparing this sample to the synthetic samples that we build. However, given that we have no magnitude information for these galaxies, they are not part of the training data for our simulation.

\subsubsection{The training set}
The $r^\prime$ and $z^\prime$ magnitudes are computed for 80,000 of the most massive galaxies in TNG300 which corresponds to most galaxies such that $\log_{10}(\mathrm{M}_{\bigstar}/(\mathrm{M}_{\odot}h^{-1}))>10$. Let us refer to the sample of all galaxies for which a set of magnitudes was computed as {\it magnitude galaxy sample}. 

As stated in the previous section, our subhalo parameters are collected from the TNG300 DM-only simulation. Due to a one-to-one matching with the hydrodynamical TNG300 simulation, we can extract corresponding baryonic parameters. Out of the 80,000 galaxies in our \textit{magnitude galaxy sample} sample, only 69,767 were successfully matched with a DM-only subhalo. The matching success rate is dependent on subhalo mass with the most massive subhaloes, where LRGs are more likely to be hosted, having a much larger probability of being matched. Approximately nine~per~cent of our \textit{true} LRGs are inside unmatched subhaloes. Three-quarters of these missing LRGs live in subhaloes smaller than 10$^{12}$\,M$_{\odot}$ where our \textit{magnitude galaxy sample} is not complete. 

While \cite{2022MNRAS.512.5793Y} attempted to match these remaining unmatched subhaloes by identifying DM-only subhaloes with a similar mass to their counterparts in the hydrodynamical simulation that inhabit a neighbouring region of the grid, we chose not to follow that approach. We made this decision because it could lead to a higher number of incorrect matches, potentially reducing the strength of the relationships between subhalo parameters and galaxy properties and potentially impacting the training of the NN.

\begin{figure}
\centering
\includegraphics[width=\linewidth]{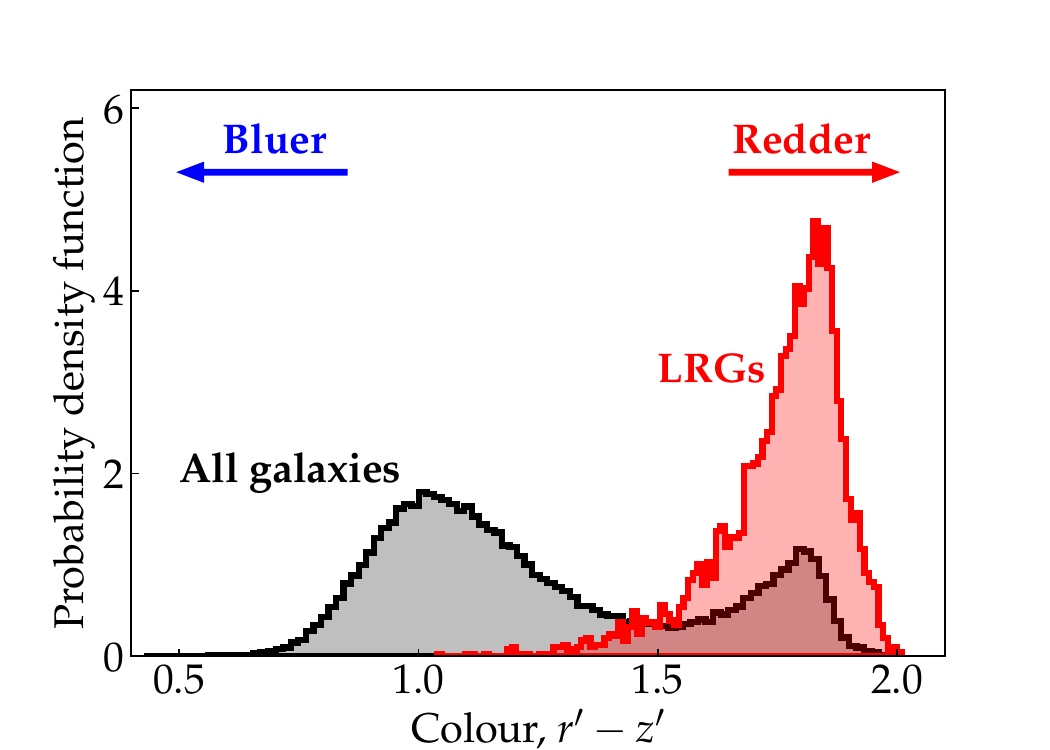}
\caption{
The distribution of galaxy $(r^{\prime}-z^{\prime})$ colours provided by \protect\cite{2022MNRAS.512.5793Y} 
for TNG300 galaxies with stellar mass greater than $10^{10} \mathrm{M}_{\odot}$ and that have a subhalo match in the DM-only and hydrodynamical versions of TNG300. The grey distribution represents all these galaxies, while red represents only those galaxies that are considered LRGs (see equation~\eqref{eq:colour_mag_cuts}). They are separately normalised to highlight that some of the LRGs have slightly bluer colours.}
\label{fig:colour_distr}
\end{figure}

Fig.~\ref{fig:colour_distr} shows the probability distribution of the colours of the matched subhaloes in the {\it magnitude galaxy sample} (grey histogram), as well as the colour distribution of matched LRGs (red histogram), where the LRGs were defined according to Section~\ref{sec:LRG_selection}. We note that the galaxy sample is bimodal with galaxies being broadly distributed into a red and a blue peak, with a boundary between both being approximately at a colour of $(r^{\prime} - z^{\prime}) = 1.5$. While our LRG sample mostly consists of galaxies in the red peak, we note that there is a small subset of LRGs ($\sim 5$~per~cent) that are bluer than this boundary. These galaxies, in general, are brighter in $z^{\prime}$ than the average LRG.

\begin{figure}
\centering
\includegraphics[width=\linewidth]{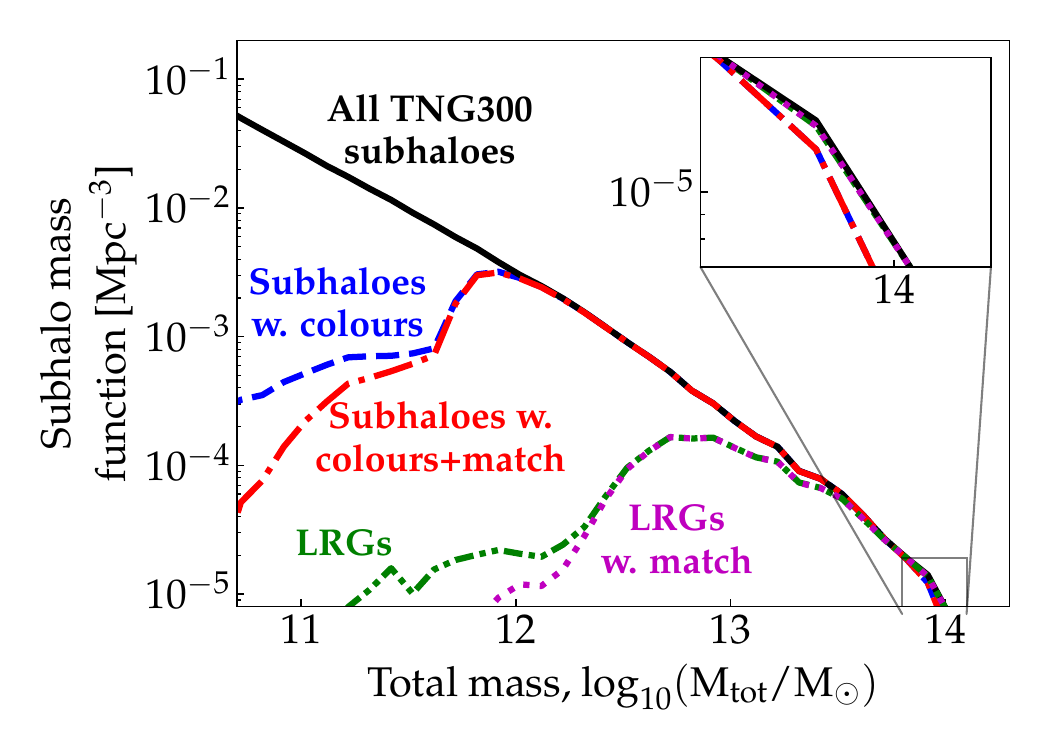}
\caption{Subhalo mass function for all subhaloes in the TNG300 (hydrodynamical) simulation (black solid line), for only the subhaloes with colours generated by \citet{2022MNRAS.512.5793Y} (dashed blue line), for the subhaloes which have both colours and a match in the DM-only simulation (dotted red line), for the LRGs (green dashed-double dotted line), and finally for the LRGs with a match in the DM-only simulation (magenta dotted line). The inset shows that there are 
a number of missing subhaloes at the high mass end. We discuss these in Appendix~\ref{sec:appendixA}.}
\label{fig:completeness}
\end{figure}

Fig.~\ref{fig:completeness} shows the mass distribution of our subhaloes, the subhalo mass function (sHMF). It compares the sHMF of matched subhaloes within our \textit{magnitude galaxy sample} (red dashed-dotted line) to the full massive galaxy sample (solid black line). The figure also shows the sHMF of all LRGs in the \textit{magnitude galaxy sample} (green dashed-double dotted line), as well as the sHMF of only those LRGs that have a match (purple dotted line).

Finally, we note that for 143 galaxies in our matched sample we cannot trace the evolution far back enough to calculate FC$_{30}$. As such our final sample that is used for the training of our NN consists of \textbf{69,625} galaxies.

\subsection{Building our LRG selector}
\label{sec:NN_train}

We have defined a sample of 69,625 galaxies to train our NN and to test its accuracy. Each galaxy in this sample has an associated value for each of the three dependent variables: $r^{\prime}$, $(r^{\prime}-z^{\prime})$, and stellar mass, as well as an individual value of all nine DM-only properties detailed in Section~\ref{sec:training_params}. Our aim is to train a NN capable of predicting the values of our dependent variables for a new DM-only subhalo based on its nine DM-only properties. In order to do so we divide our galaxy sample into three parts. Sixty~per~cent is randomly selected as our training set, twenty~per~cent is selected as our validation set and is used to assess the accuracy of the model after each training epoch. The remaining twenty~per~cent constitutes our test set which is kept separate from the training and validation process and is used to evaluate the model's accuracy. This prevents unwarranted trust in the accuracy of the model in the case of over-fitting. 

As detailed in Section~\ref{LRGS_selection_NN}, our validation set is also used to update the selection criteria of equation~\eqref{eq:colour_mag_cuts} so that it selects the right number density of LRG subhaloes using the values predicted by our NN.

When running a NN, it is convenient to keep all neuron inputs and outputs within a similar range of values. With this in mind, if $\vec{\theta}$ is the set of values from all galaxies of either one of the DM only or dependent parameters, and $\theta_{i}$ is the value of that parameter for the $i$-th galaxy, we define the normalized parameters $\bar{\theta}$ as \ $\bar{\theta}_{i} = [\theta_{i} - \mu(\vec{\theta})]/\sigma(\vec{\theta})$, where $\mu$ and $\sigma$ are the mean and standard deviation operators, respectively. This is true for all parameters except the central/satellite flag, which is set to 1 for centrals and -1 for satellites.

Our NN architecture employs a Multi-Layer Perceptron, comprising of two hidden layers, each consisting of 100 neurons. We adopt the activation function suggested by \cite{2020ApJS..249....5A} and \cite{2022JCAP...04..056D}:

\begin{equation}
    a(X)=\left[\gamma+\frac{1-\gamma}{1+e^{-\beta X} }\right] X,
\end{equation}

where $\gamma$ and $\beta$ are new free parameters of each neuron in each hidden layer which are fitted during the training of the NN. Given that we have two hidden layers with 100 elements each, this adds 400 parameters to the fit. A NN of this size has tens of thousands of free parameters, therefore these extra 400 parameters do not increase the work of the NN in a significant manner, while they allow the activation function to reproduce smooth or sharp changes in gradients better, as discussed in \cite{2020ApJS..249....5A}. We find that this activation function leads to more accurate models and requires fewer training epochs to reach the final solution when compared to other commonly used activation functions, like Rectified Linear Units. This is in agreement with what, e.g.\ \cite{2020ApJS..249....5A} and \cite{2022JCAP...04..056D} have found. At each epoch, we monitor the mean square error (MSE) of the validation set. We keep track of the smallest MSE found so far and of how many epochs have passed since it was found: if no improvement is seen for 100 epochs we consider our training complete.

\subsubsection{NN results}

\begin{figure}
\centering
\includegraphics[width=\linewidth]{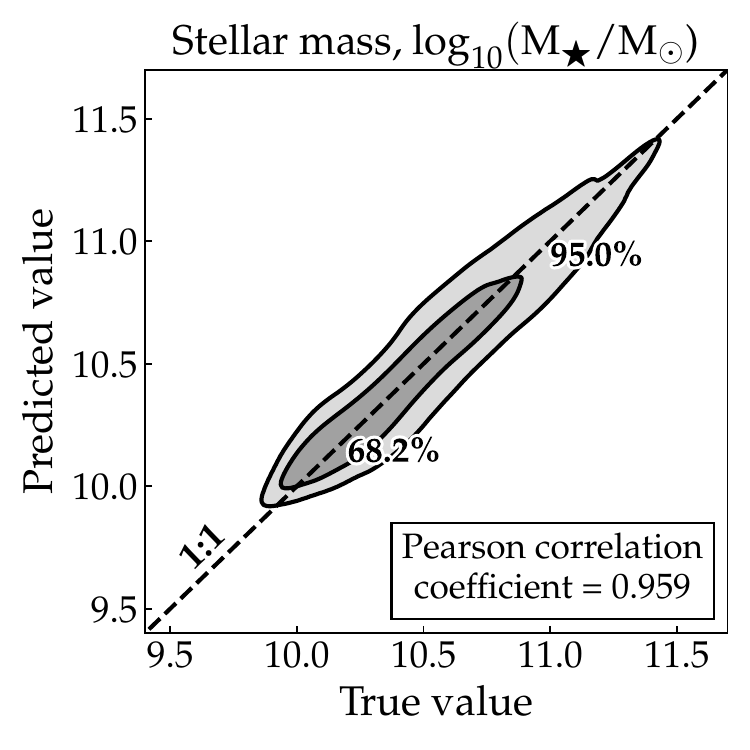}
\caption{Density contours of the predicted versus true values for the (log of the) stellar mass of the \textit{test} sample. The dashed black line (labelled 1:1) indicates where the two are equal. The Pearson correlation coefficient is indicated in the key and measures the linear correlation between the two distributions.}
\label{fig:pred_mstar}
\end{figure}

\begin{figure}
\centering
\includegraphics[width=\linewidth]{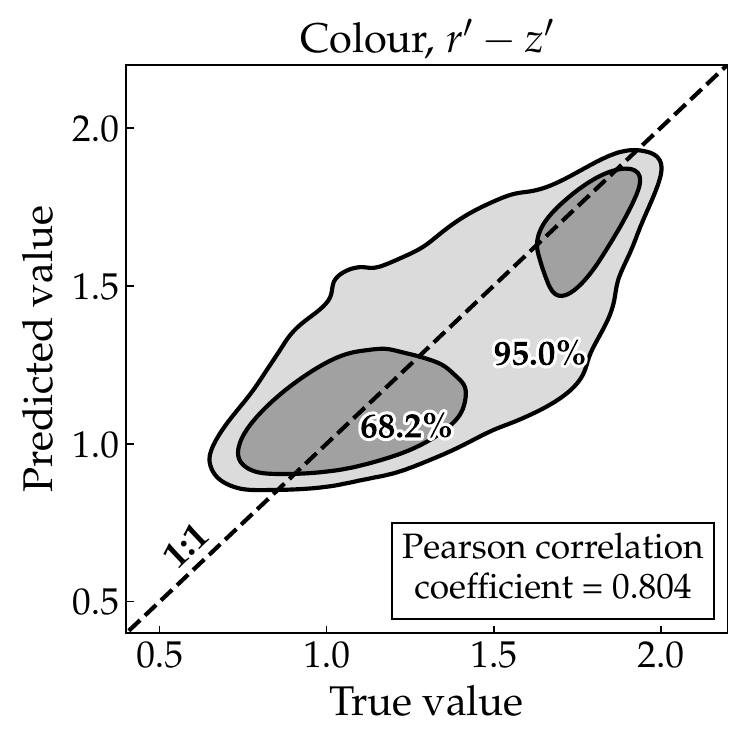}
\caption{Similar to Fig.~\ref{fig:pred_mstar}, but for predicted versus true $(r^{\prime}-z^{\prime})$ colour. The bimodality of the colour distribution is preserved in the predicted $(r^{\prime}-z^{\prime})$ colour, even though not perfectly.}
\label{fig:pred_colour}
\end{figure}

\begin{figure}
\centering
\includegraphics[width=\linewidth]{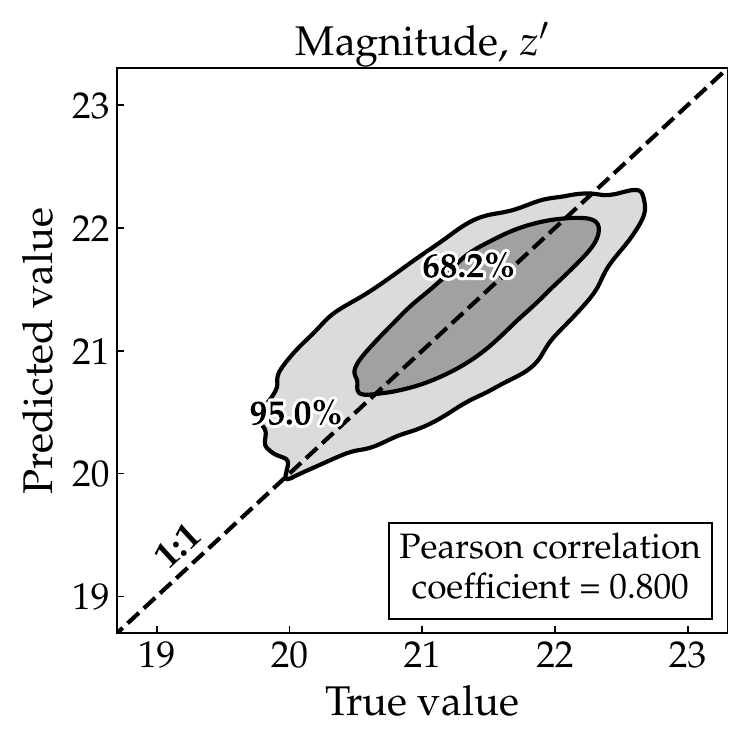}
\caption{Similar to Fig.~\ref{fig:pred_mstar} and \ref{fig:pred_colour}, but for predicted versus true $z^{\prime}$ magnitude.}
\label{fig:pred_mag}
\end{figure}

We assess the accuracy of our neural networks (NNs) in predicting the values of our dependent variables using the test set introduced in Section~\ref{sec:NN_train}. This test set is excluded from the model's training process, allowing us to evaluate performance without the risk of overfitting biasing our conclusions. Fig.~\ref{fig:pred_mstar}, \ref{fig:pred_colour}, and \ref{fig:pred_mag} show the predictions of the three dependent variables compared with the true values of the variables for all points in our test set. The black dashed line in each figure corresponds to the one-to-one relation and the closer a galaxy is to this line, the better the prediction of the model is. 

A common statistical tool to quantify the accuracy of a predictive model is the Pearson correlation coefficient, $\pcc$. It measures the linear correlation between two samples and are normalized in such a way that for a perfect correlation, where all galaxies lay exactly on top of the one-to-one line, $\pcc=1$, while for two completely uncorrelated variables $\pcc=0$. In each figure, the Pearson correlation coefficients are indicated in the key, with stellar mass being by far the most accurately predicted with $\pcc=0.959$ and our ($r^{\prime}-z^{\prime}$) and $z^{\prime}$ models measuring $\pcc=0.804$ and $\pcc=0.800$ respectively.

To asses the performance of our NN we compare these $\pcc$ values to those found by other groups. Many other teams have build ML models to predict the stellar mass of the TNG300 galaxies as well as other hydrodynamical simulations. In general those works find $\pcc$ values between 0.92 to 0.98 for their stellar mass models \citep[e.g.][]{Agarwal_2018,Lovell_2022,de_Santi_2022}, which is consistent with our value of $\pcc=0.96$. 

It is more challenging to make confident claims about the effectiveness of our model in capturing the other two dependent properties, as to our knowledge, there are no other ML models trained on the magnitude dataset from \cite{2022MNRAS.512.5793Y}. Hence a direct comparison is not possible. However, other ML approaches have modelled colours computed from the TNG300 AB absolute magnitudes presented in \cite{Nelson_2017}. Note that our observer frame magnitudes differ from this rest-frame magnitudes in many aspects.

Nevertheless, we find that the $\pcc$ of our predictions for $(r^{\prime}-z^{\prime})$ colours are of a similar order of magnitude as the $(g-r)$ predictions reported in \cite{de_Santi_2022} and \cite{Chuang_2023}, which achieved coefficients as high as $\pcc=0.7$. Although we achieve a higher $\pcc$ value for $(r^{\prime}-z^{\prime})$ of $\pcc=0.81$, it is worth noting that modelling bluer colours is inherently more challenging due to the additional complexity introduced by star formation processes.

The differences in the properties modelled, coupled with substantial variations in methodology, pose challenges in precisely quantifying the performance of our model in colour selection. However, the relatively similar $\pcc$ values offer some reassurance regarding the consistency of our results.

Another notable observation revealed by Fig.~\ref{fig:pred_mstar}, \ref{fig:pred_colour}, and \ref{fig:pred_mag} is the asymmetry of the distribution around the one-to-one line. Fig.~\ref{fig:pred_mag} demonstrates that, generally, our model tends to underestimate $z^{\prime}$ for faint {\it true} galaxy magnitudes, while it tends to overestimate brighter {\it true} galaxy magnitudes. In other words, our model is more inclined to predict brighter galaxies as slightly fainter than they truly are and vice-versa. This is also visible in Fig.~\ref{fig:pred_mstar} where at very large stellar masses there is more coloured area bellow the one-to-one line than above it. This indicates that our model is more likely to under-predict the $\mstar$ associated with a subhalo than to over predict it. Similar asymmetries are also present in our $(r^{\prime}-z^{\prime})$ predictions where our model is more likely to predict red galaxies to be slightly redder than what they actually are (although this is slightly less obvious to spot by eye in Fig.~\ref{fig:pred_colour}).

As mentioned earlier, we identify LRGs by setting threshold values within our parameter space and selecting all galaxies falling within these thresholds as LRGs. The asymmetry observed above will impact the density of selected LRGs. Given our tendency to under-predict values, there is a higher likelihood of incorrectly classifying a galaxy as too blue or too dim to be an LRG than mistakenly selecting a bluer and brighter galaxy as an LRG. Bearing this in mind, we slightly adjust our selection criteria when identifying LRGs in our predicted parameter space to maintain a consistent density of LRGs.

\subsubsection{Selecting LRGs from our NN model predictions}
\label{LRGS_selection_NN}

DESI LRGs at $z\sim0.8$ are characterised as galaxies which magnitude and colours satisfy equation~\eqref{eq:colour_mag_cuts}. Due to the asymmetries of the predictions in the properties around the one-to-one line as discussed in the previous section, the density of LRGs larger than a given cut will be different when applied to our {\it true} and predicted dependent variables. Therefore if the cuts of equation~\eqref{eq:colour_mag_cuts} remain unchanged, the LRG density of our mock sample will be different to that expected from DESI data. Therefore we modify slightly the LRG cut, in such a way that we keep the number density of LRGs unchanged while also selecting as many {\it true} LRGs as possible.

We have developed NN models for three distinct galaxy parameters. In what follows, we investigate how various cuts within these predicted parameter spaces can effectively segregate samples into LRGs and other galaxies. We aim to find cuts that conserve the number density of LRGs while maximizing the identification of true LRGs. We explore two different approaches: one by making a cut in the colour-stellar mass space and the other in the colour-magnitude space. In the former, we aim to select red and massive galaxies as LRGs, while in the latter, we focus on selecting red and bright galaxies as LRGs. 

We note that is not straightforward which approach is more optimal. The colour-magnitude space is where the original DESI cuts we seek to emulate were made, whereas stellar mass $\mstar$ is modelled much more accurately by our NN models than the $z^{\prime}$-band magnitude.

The new cuts are determined as follows. We use the Python function \textsc{scipy.optimize.minimize} with the Powell algorithm \citep{2002nrca.book.....P} to fit a straight line to the colour versus stellar mass/magnitude predictions by the NNs of the \textit{validation} set to select LRGs. To select the best cut, we simultaneously minimize the difference in the number of true and predicted LRGs, and (one minus) the $F_{1}$ score. The $F_{1}$ score is a measure of predictive performance, and can be thought of as the harmonic mean of the precision and recall, where precision is the number of true predicted LRGs divided by the number of all galaxies predicted to be an LRG, and recall is the number of true predicted LRGs divided by the number of all galaxies that should have been identified as an LRG. The $F_{1}$ score ranges from 0 to 1 with 1 indicating perfect precision and recall.

With the method described above, the number of predicted LRGs is fixed to be precisely the same as the number of true LRGs in our validation set. These cuts, determined on the \textit{validation} set, are then used to separate galaxies from the \textit{test} set into LRGs and regular galaxies.

\begin{figure}
\centering
\includegraphics[width=\linewidth]{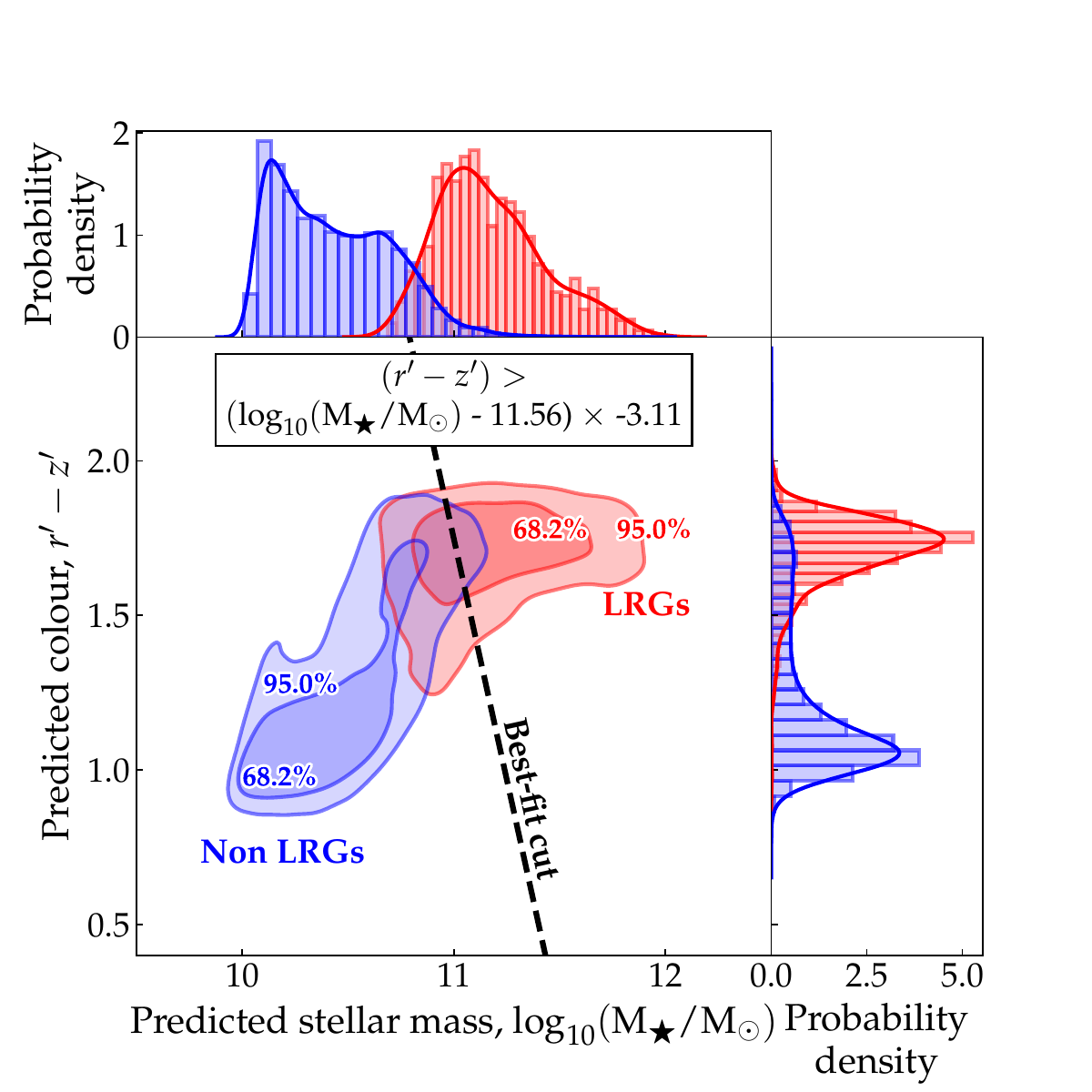}
\caption{Predicted colour ($r^{\prime}-z^{\prime}$) plotted against the predicted (log of the) stellar mass for the \textit{test} set. Galaxies classified as an LRG using equation~\eqref{eq:colour_mag_cuts} are shown in red, all other galaxies in blue. The contours indicate 68.2 and 95 percentiles of the distribution in the plane for LRGs and non LRGs respectively, and were calculated using kernel density estimation (KDE). The top and right hand panel show histograms of predicted stellar mass and colour for both LRGs and non LRGs with 1D KDEs overlaid (solid lines). The best-fitting cut, indicated in the key, is shown as a black dashed line. With this cut, 69.06 and 98.19~per~cent of the LRGs and non LRGs respectively are classified correctly.}
\label{fig:SM_r_z_cut}
\end{figure}

\begin{figure}
\centering
\includegraphics[width=\linewidth]{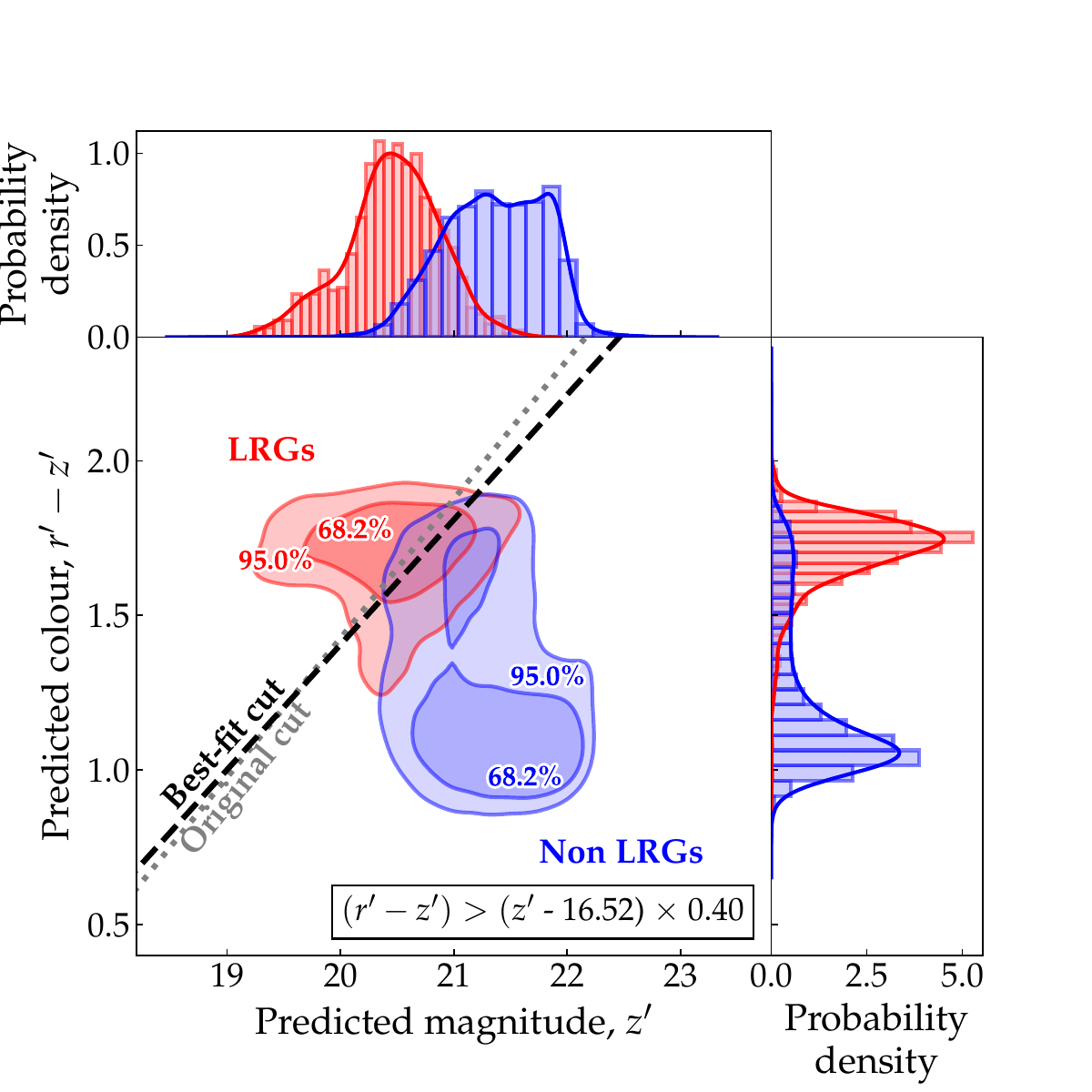}
\caption{Predicted colour ($r^{\prime}-z^{\prime}$) plotted against the predicted $z^{\prime}$ magnitude for the \textit{test} set. Line styles and panel layout are similar to that in Fig.~\ref{fig:SM_r_z_cut}. The nominal colour-magnitude cut from equation~\eqref{eq:colour_mag_cuts} is shown as a dotted grey line, while the best-fitting cut, indicated in the key, as a black dashed line. With this improved cut, 68.58 and 98.07~per~cent of the LRGs and non LRGs respectively are classified correctly.}
\label{fig:z_r_z_cut}
\end{figure}

Fig.~\ref{fig:SM_r_z_cut} and \ref{fig:z_r_z_cut} show the predictions for colour, $(r^{\prime}-z^{\prime})$, versus stellar mass and $z^{\prime}$-band magnitude respectively for galaxies in the \textit{test} set. They also show the equations utilized throughout this work to define our cuts in both the colour-stellar mass plane and the colour-magnitude plane respectively.

The colour-stellar mass plane and colour-magnitude plane cuts select 815 and 826 LRGs respectively, which is similar to 837, the \textit{true} number of LRGs in the \textit{test} set 
 (considering the fits were performed to the \textit{validation} set). 

The number of LRGs is much smaller than that of regular galaxies. Consequently, relatively small uncertainties in the predicted properties of non LRG galaxies translate into a large number of false-positive galaxies selected as LRGs. This means that in order to conserve the density of galaxies, our cuts prioritize the purity of the sample (reducing the number of false positives) over the completeness of the sample (reducing the number of false negatives). As a consequence, the colour-stellar mass plane and colour-magnitude plane cuts select 69.06~per~cent and 68.58~per~cent of true positive LRGs, with the remainders corresponding to false-positive non LRG galaxies that make it into our sample.

Through the rest of this work we utilize the stellar mass and colour cut relation shown in Fig.~\ref{fig:SM_r_z_cut} to select LRGs in the predictions of our NN models. We note that similar results are obtained if one uses the magnitude and colour cut relation shown Fig.~\ref{fig:z_r_z_cut}.

Finally, we note that our methodology for selecting LRGs has some intrinsic variation due to the randomness involved when dividing our data into training, test, and validation sets. To test the effect of this variation we ran our NN four times with different initial seeds for our training, test, and validation sets. This is explored in detail Appendix~\ref{sec:appendixB}.

\subsubsection{Testing the resulting LRG sample}

In what follows, we quantify the similarity of the sample of LRGs selected using the cuts presented in the previous section in comparison to our sample of {\it true} galaxies. Throughout this work, we refer to the sample constructed by populating our test set with subhaloes as our TNG300 sample, distinguishing it from our {\it true} LRG sample.

\begin{figure}
\centering
\includegraphics[width=\linewidth]{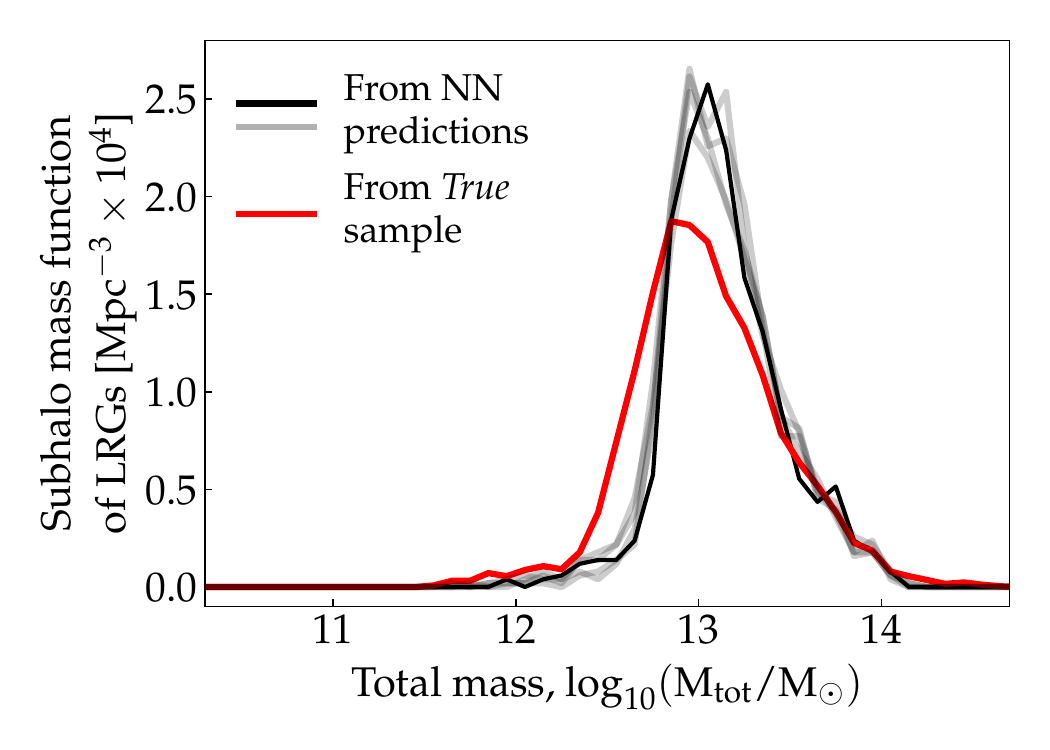}
\caption{The (sub)halo mass function of the \textit{true} LRG sample is represented by the red line, while the predicted TNG300 LRG sample is shown by the black line. The latter was derived by applying the selection criteria outlined in Fig.~\ref{fig:SM_r_z_cut} to predict the colours and magnitudes of our test set. The four faint black lines correspond to different TNG300 LRG samples obtained by re-running our entire methodology with different random seeds for dividing our data into training, test, and validation subsets. All samples are incomplete, as they do not include LRGs within unmatched subhaloes.}
\label{fig:HMF}
\end{figure}

Fig.~\ref{fig:HMF} illustrates the halo mass function (HMF) of both the {\it true} LRG sample (red) and our predicted LRG TNG300 sample (black), the halo mass function is defined as the density of subhaloes of a given mass per unit volume.

The {\it true} curve encompasses all our 4234 matched LRGs, including the 42 more massive LRGs for which we have no colour, but excluding those unmatched LRGs for which we do not have an associated subhalo mass. The black TNG300 sample comprises of only the LRGs from our test sets. The TNG300 has approximately five times less LRGs than our {\it true} sample and therefore a difference in shot noise between both curves should be present. 

Given that our test set is approximately five times smaller, we normalize the volume of the simulation  to be approximately one-fifth of its true value. After this normalization, the number density $N_{\rho}$ of LRGs in each sample is nearly identical, with the {\it true} LRG sample having a density of $N_{\rho}=1.61 \times 10^{-4}\,\mathrm{Mpc}^{-3}$ and our predicted LRG sample having a density of $N_{\rho}=1.67 \times 10^{-4}\,\mathrm{Mpc}^{-3}$. Given that the densities are so similar, the area under each curve is almost identical for both samples.

The plot shows that our model encounters some difficulty in selecting smaller host haloes, as evidenced by the red curve being above the black lines for $\log_{10}(\mathrm{M_{tot}/M_{\odot}}) < 12.7$. Given that the density of galaxies is similar, this feature makes the red curve wider and shorter. Consequently, our model tends to over predict the number of LRGs in haloes around $\log_{10}(\mathrm{M_{tot}/M_{\odot}}) = 13$, where the black curve peaks.

The inability to select smaller haloes might be due to missing information from the DM-only halo properties that we use to build our model. To understand this limitation, it is important to note that our NN is sensitive to certain galaxy evolution processes that partially determine the colour and brightness of galaxies. Our subhalo parameters effectively track the age and size of host subhaloes, and generally, older haloes become redder while massive haloes are more luminous. However, additional processes driven by baryons, and therefore absent in DM-only simulations, can influence galaxy evolution and may contain information missing from our models. For instance, AGN feedback is well-known to play a significant role in quenching galaxies, which results in reduced star formation and therefore redder galaxies.

Since our DM-only informed NN lacks information on AGN feedback and other baryonic processes, it cannot account for these effects. Therefore, our model tends to select large and old subhaloes as LRGs because they have grown through the accretion of matter over long periods and multiple mergers. However, it misses smaller subhaloes that might turn red or bright due to these baryonic processes.

The fraction of satellite subhaloes hosting LRGs is around 14.9~per~cent in the {\it true} sample and around 14.3~per~cent in the TNG300 sample. These numbers are quite similar, especially considering that the TNG300 sample selects larger host haloes, which are more likely to be central haloes. Both of these numbers are in agreement with the expected percentages from DESI LRGs. 

\section{Selecting LRGs in the MultiDark Planck 2 simulation}\label{sec:data_MDPL2}

We have introduced a methodology to build a NN model capable of selecting DESI-like LRG host subhaloes, and we discussed how the resulting distribution of LRGs compares to our original sample. Next, we will use this model to populate a large 1\,(Gpc\,$h^{-1}$)$^{3}$ simulation with LRGs and analyse how the resulting LRG sample statistics compare to those obtained from our smaller TNG300 box model.

We begin in Section~\ref{MDPL2_intro} by introducing the large box simulation we use and listing all relevant features. Then, in Section~\ref{sec:MDPL2_sample}, we describe the procedure we follow to extract the nine halo parameters that our model needs. Here, we also compare the distribution of these parameters between both simulations. Finally, in Section~\ref{sec:sec_MDPL2_lrgs}, we utilize our NN model to populate the large box simulation with LRGs.

\subsection{The MDPL2 Simulation}
\label{MDPL2_intro}

The MultiDark Planck 2 simulation \citep[MDPL2, ][]{Klypin_2016} is a large DM-only N-body simulation with a volume of 1\,(Gpc\,$h^{-1}$)$^{3}$ and a DM particle mass of $1.5\times10^{9}$\,M$_{\odot}\,h^{-1}$. The MDPL2 simulations were carried out with \textsc{L-gadget-2}, a modified version of the cosmological code \textsc{gadget-2} described in \citet{Springel_2005b}. MDPL2 is built under a $\Lambda$CDM paradigm with the cosmological parameters of \citet{Planck_2013}: $\Omega_\mathrm{m} = 0.307115$, $\Omega_\mathrm{b} = 0.048206$, $h = 0.6777$, $n_{s} = 0.96$, and $\sigma_{8} = 0.8228$. The outputs of the simulations are saved in 126 redshift snapshots ranging from starting redshift $z = 17$ to $z = 0$. MDPL2 has the same box size, cosmological parameters and particle resolution as the MDPL simulation, but has a different initial seed. We use MDPL2, since there are no merger trees available for MDPL, which we require for our modelling. We use snapshot 98 of MDPL2 which has a redshift of $z=0.82$.

The \textsc{Rockstar} \citep{Behroozi_2013a} halo-finder code is used to identify haloes within the simulation using a FoF approach. As with TNG300, a linking length of $b = 2$ is used. \textsc{Rockstar} clumps substructures in both position and velocity space to identify subhaloes within each halo.Then, merger trees are built linking each (sub)halo with its progenitors with the \textsc{consistent trees} procedure of \citet{Behroozi_2013b}.

We define the total subhalo mass in MDPL2 to be M$_\mathrm{vir}$, the mass enclosed within the virial radius\footnote{In the case of MDPL2, this corresponds to the radius within which the average density is 360 times the background density of the Universe at redshift $z = 0$.}. In TNG300, the total subhalo mass is the sum of all particles gravitationally bound to the subhalo. We account for the different mass definitions by including a shift of 0.082\,dex in the log of the total mass in MDPL2. We discuss this in more detail in Appendix~\ref{sec:appendixA}.

\subsection{Our MDPL2 halo sample}\label{sec:MDPL2_sample}

There are 140 million subhaloes within the $z=0.82$ redshift slice of MDPL2. Given the large number of subhaloes, tracking their evolution and computing the formation criteria parameters for all of them becomes computationally expensive. However, this large cost is unnecessary since we know that LRGs only reside inside subhaloes that are massive or have been massive at some point during their evolution. With this in mind, we only compute the formation criteria parameters for subhaloes that satisfy $\log_{10}(\mathrm{M_{max}/M_{\odot}}) > 10.95$, which correspond to the most massive ${\sim}55$ million subhaloes. Only subhaloes that meet this criteria are populated with galaxies using our NN. This cut should be sufficient and include all reasonable LRG candidates, considering that all LRGs in the TNG300 dataset are inside much larger subhaloes with our smaller LRGs being inside host subhaloes of around $\log_{10}(\mathrm{M_{max}/M_{\odot}}) = 11.7$. 

We utilize the same algorithm from Section~\ref{sec:training_params} to extract the evolution parameters. We first smooth the formation history of the halo mass with a Gaussian kernel and then compute M$_\mathrm{max}$, the redshift at which the maximum subhalo mass is reached, and the formation criteria parameters from the resulting smoothed curve. We note that the TNG300 and MDPL2 do not have snapshots at exactly the same redshifts. We therefore only select the snapshots in each simulations that are within 0.05 of each other. Additionally, we select around one-third of these snapshots. The reason for selecting this small subset is twofold. First, we are only selecting snapshots that have a TNG300 counterpart at a similar redshift, ensuring that the evolution of each galaxy is traced on equivalent slices between both simulations. Second, each slice requires significant computational memory to store and load. By reducing the number of slices, we keep the computational cost more manageable.

The resolution of MDPL2 is lower than that of TNG300, resulting in a smaller number of subhaloes for which the \textsc{Rockstar} algorithm can find a progenitor with 30~per~cent of the maximum mass. For our cut-off of $\log_{10}(\mathrm{M_{max}/M_{\odot}}) > 10.95$, this progenitor would consist of around 12 particles. This is not a lot, and there is a chance the algorithm might not identify it as an independent subhalo. This issue affects approximately 10~per~cent of the MDPL2 haloes above our mass cut.

For these subhaloes, we replace FC$_{30}$ with the earliest redshift at which a progenitor is detected. For the vast majority of these subhaloes, this early progenitor has a mass that is only slightly more than 30~per~cent of its maximum mass. While this is not the ideal procedure, we note that only around 0.5~per~cent of our selected LRGs have an ill-defined FC$_{30}$. Thus, we consider this effect to be minimal, especially since all other eight parameters are well defined.

\begin{figure*}
\includegraphics[width=0.95\textwidth]{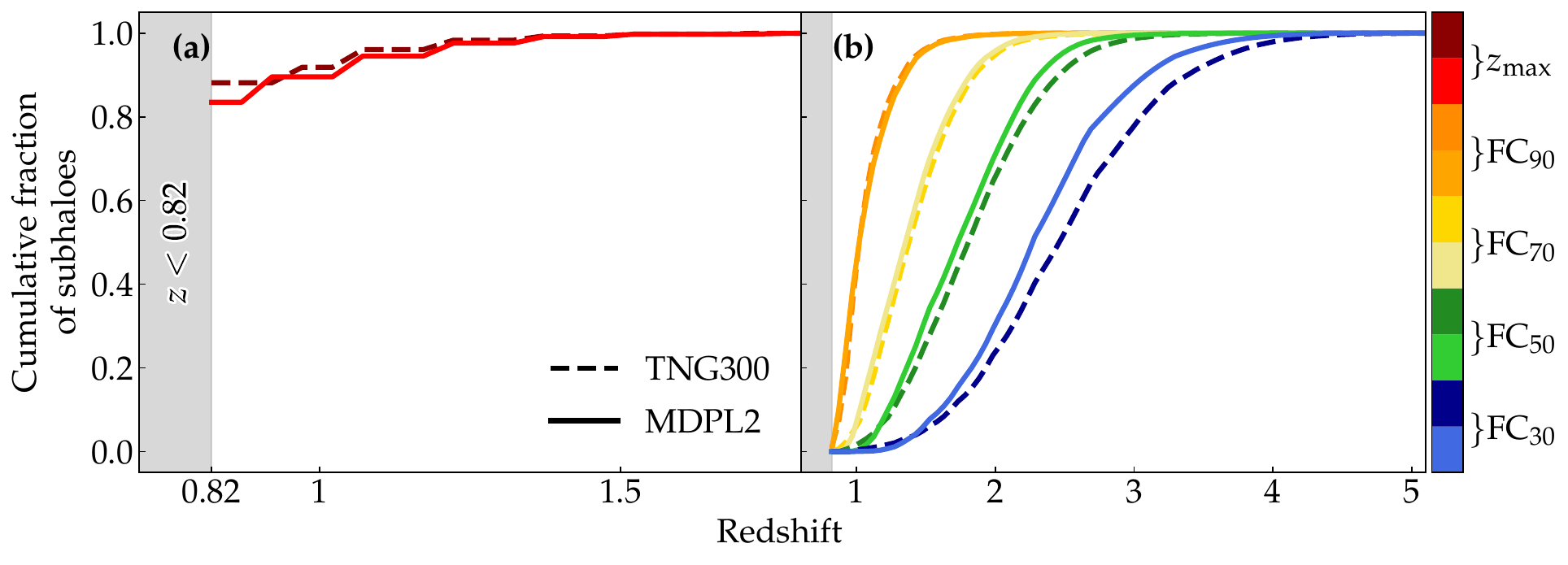}
\caption{Cumulative histograms of the formation criteria outlined in Section~\ref{sec:training_params}: the redshift at which the maximum mass was formed, $z_\mathrm{max}$ (a), and the redshifts at which 90~per~cent, 70~per~cent, 50~per~cent, and 30~per~cent of the maximum mass was formed, FC$_{90}$, FC$_{70}$, FC$_{50}$, and FC$_{30}$ respectively (b). For a given parameter, the y-axis value represents the fraction of subhaloes with a value of that parameter \textit{smaller} than the corresponding x-axis value. I.e.\ in (b), for about half of the subhaloes FC$_{90}$ (orange solid and dashed lines) is equal or smaller than a redshift of 1.}
\label{fig:evo_hists}
\end{figure*}

\begin{figure*}
\includegraphics[width=0.95\textwidth]{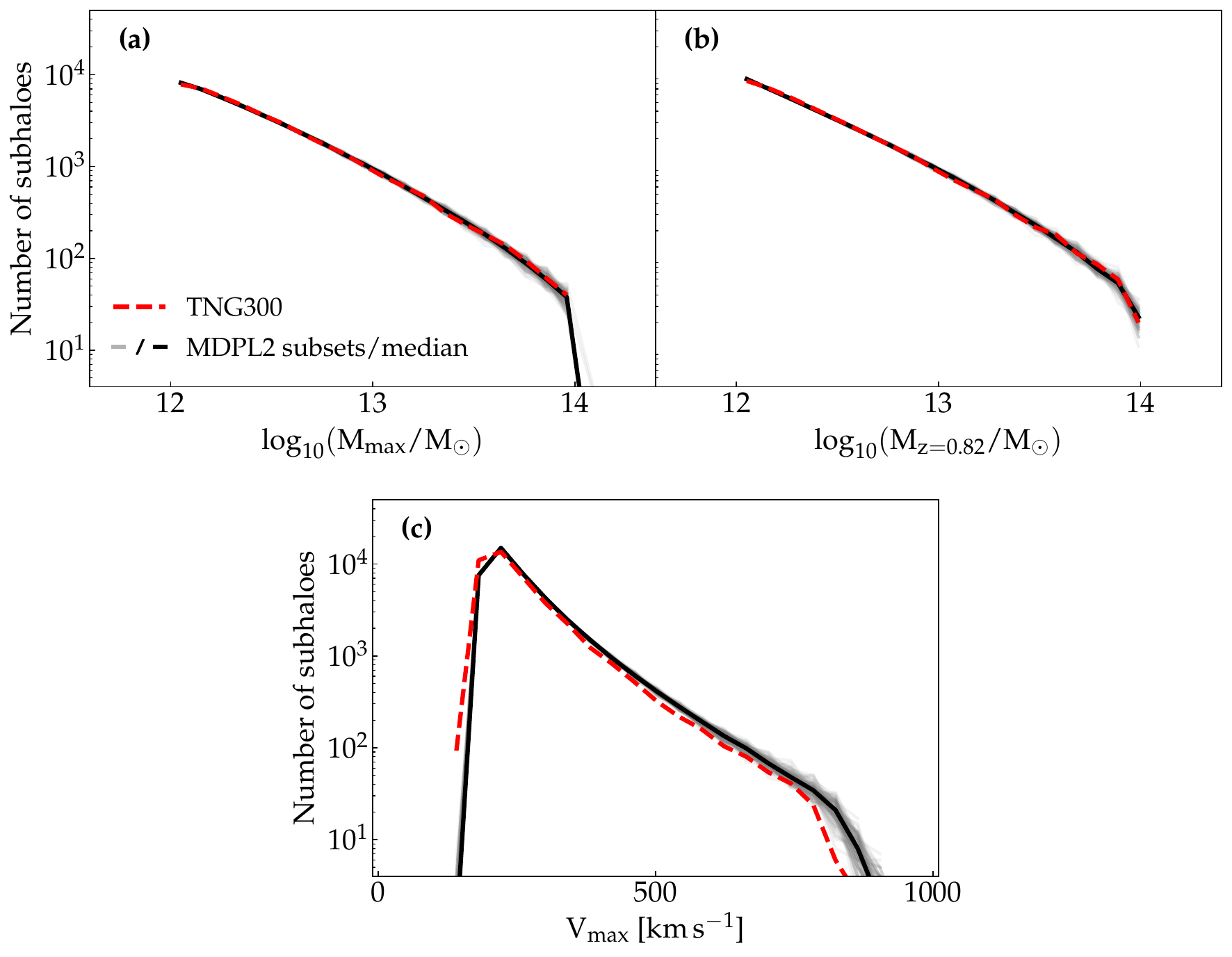}
\caption{Histograms comparing the distribution of three of the training parameters for the NN outlined in Section~\ref{sec:training_params}: the maximum subhalo mass along its formation history up until and including $z=0.82$, M$_\mathrm{max}$ (a), the total subhalo mass at $z=0.82$ M$_{z=0.82}$ (b), and the maximum circular velocity of the subhalo at redshift $z=0.82$ V$_\mathrm{max}$ (c). The red dashed line represents TNG300, while the grey solid lines represent 100 random sub-samples from MDPL2 that have the same size as the TNG300 sample, with the solid black line being the median of these 100 lines. The distributions of plot (a) and (b) are similar due to the fact that the maximum mass (a) is most likely to be the mass at the lowest redshift, i.e.\ at $z = 0.82$ (b).}
\label{fig:mass_hists}
\end{figure*}

Fig.~\ref{fig:evo_hists} and \ref{fig:mass_hists} show the distribution of the training parameters outlined in Section~\ref{sec:training_params} for our {\it magnitude galaxy sample} from the TNG300 DM-only simulation, and compares this to the distribution of the parameters in MDPL2. The figures are constructed for all subhaloes that satisfy $\log_{10}(\mathrm{M_{z=0.82}/M_{\odot}}) > 12$, which is the mass threshold at which our {\it massive} galaxies become complete, as demonstrated in Fig.~\ref{fig:completeness}. Additionally, the plot only includes subhaloes that satisfy $\log_{10}(\mathrm{M_{z=0.82}/M_{\odot}}) < 14$, as the TNG300 simulation does not have sufficient volume to accurately compute the density of subhaloes of this size.

Fig.~\ref{fig:evo_hists} shows the distribution of our formation criteria $z_\mathrm{max}$ (a), FC$_{90}$, FC$_{70}$, FC$_{50}$, and FC$_{30}$ (b) as cumulative (fractional) histograms. The dashed lines represent TNG300, the (lighter) solid lines MDPL2. The different colours refer to the different criteria (see the colour bar). In general, we observe that the distributions are quite similar, and for the first three parameters, our TNG300 distributions are almost indistinguishable from an MDPL2 subset, with FC$_{50}$ and FC$_{30}$ being the exceptions. TNG300 and MDPL2 have do not have the same mass resolution, which could contribute to the small differences in our distributions. For a subset of the MDPL2 subhaloes, the progenitors with 30~per~cent and 50~per~cent of the mass are formed by a small number of particles. Therefore, their galaxy history would have a larger resolution uncertainty and be more affected by specific peculiarities of the procedure, such as how particular particles are selected to be (or not be) subhalo members, or the smoothing of the formation history.

Fig.~\ref{fig:mass_hists} shows the distribution of the three remaining training parameters, M$_\mathrm{max}$ (a), M$_{z=0.82}$ (b), and V$_\mathrm{max}$ (c), as ordinary (non cumulative) histograms. The red dashed line represents TNG300, the grey lines correspond to 100 random subsets of MDPL2 with an equal number of galaxies as the TNG300 sample, and the black solid line represents the median of these MDPL2 subsets.

We find that the distribution of M$_\mathrm{max}$, as well as of M$_{z=0.82}$, in TNG300 and MDPL2 are extremely similar. The distributions of V$_\mathrm{max}$ are relatively similar at the lower end, but start to differentiate more clearly at ${\sim}$750\,km\,s$^{-s}$.

\subsection{MDPL2 LRGs}
\label{sec:sec_MDPL2_lrgs}
We have shown that the distribution of parameters within our MDPL2 sample closely resembles those from our hydrodynamical simulation. With this in mind, we employ our NN models to predict the stellar mass, the $(r^{\prime}-z^{\prime})$ colour, and the $z^{\prime}$ magnitude of all the MDPL2 subhaloes above our mass cut. These predictions are then utilized to select an LRG MDPL2 sample, wherein subhaloes are chosen based on whether their predicted quantities satisfy the inequality outlined in the label of Fig.~\ref{fig:SM_r_z_cut}.

\begin{figure}
\centering
\includegraphics[width=\linewidth]{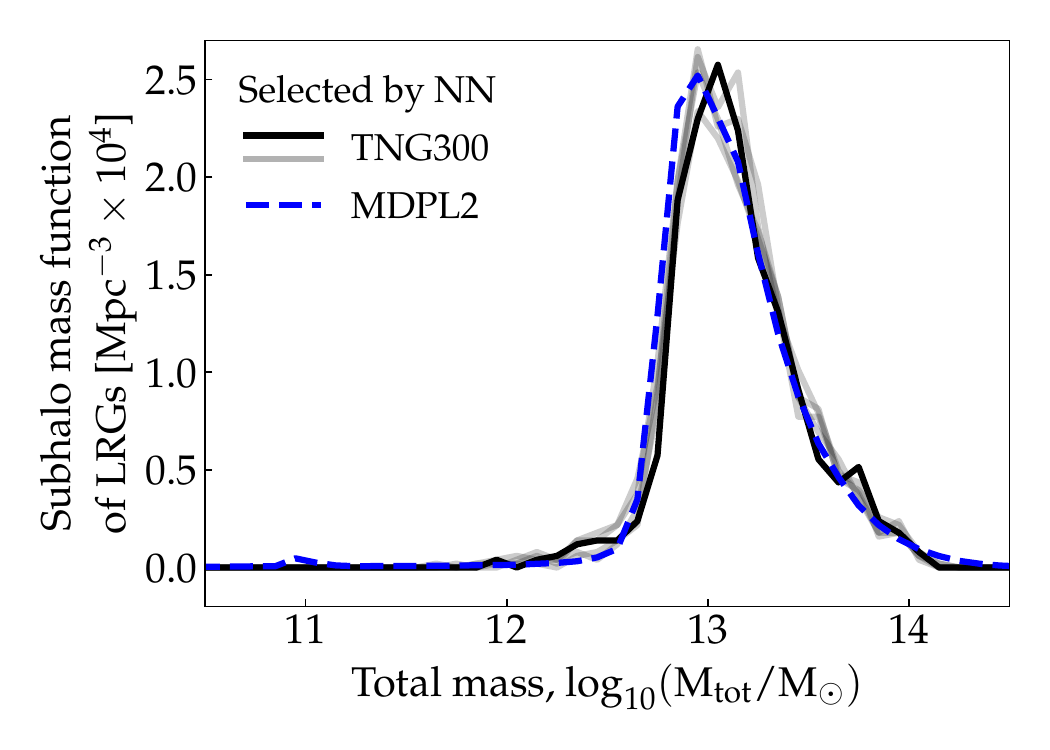}
\caption{The subhalo mass function of the predicted MDPL2 sample (blue dashed line), and of the predicted TNG300 LRG sample (solid black line) from our fiducial NN. The latter was derived by applying the selection criteria outlined in Fig.~\ref{fig:SM_r_z_cut} to predict the colours and magnitudes of our test set. The four faint black lines correspond to different TNG300 LRG samples obtained by re-running our entire methodology with different random seeds for dividing our data into training, test, and validation subsets (see also Appendix~\ref{sec:appendixA}). All samples are incomplete, as they do not include LRGs within unmatched subhaloes. Note that the solid black and grey lines are identical to those of Fig.~\ref{fig:HMF}.}
\label{fig:HMF2}
\end{figure}

Fig.~\ref{fig:HMF2} shows the sHMF of the MDPL2 sample and of our fiducial TNG300 sample (and of the four test NN runs). We find that these sHMFs are similar and that the subhalo distributions of both samples are comparable. Both samples also show relatively similar densities of LRGs, with the MDPL2 sample containing 1.5~per~cent more LRGs than what is expected from the DESI LRG density. Note that the sHMF from TNG300 only contains LRG predictions within our TNG300 test set. There should therefore be a difference in shot noise between the two curves. 

We also note that all subhaloes that satisfy $\log_{10}(\mathrm{M_\mathrm{vir}/M_{\odot}})>13.8$ in our MDPL2 LRG sample are selected as LRGs, which reassures us that even if our training set is missing some of these larger galaxies due to the stellar mass upper cut in our {\it magnitude galaxy sample}, our NN model in MDPL2 is consistent with our decision of including all these missing galaxies into our \textit{true} TNG300 LRG sample. 

The fraction of central and satellite subhaloes populated with LRGs is different, however, in our MDPL2 sample where the fraction of satellites is 10~per~cent. Which we acknowledge is lower than the $\sim 14$~per~cent from the TNG300 sample. While the exact reason for this discrepancy is unclear to us, it may be related to intrinsic differences between the TNG and MDPL2 simulations. Notably, we observe a difference in the overall central-to-satellite ratio between the two simulations, with ${\sim}$83~per~cent of haloes in the relevant mass range being centrals in MDPL2, compared to ${\sim}$85~per~cent in TNG300.

\section{Evaluating the HOD Galaxy-Halo Connection}
\label{section4}
We have constructed our NN LRG galaxy sample within the MDPL2 simulation using our NN predictions. This methodology aims to select galaxies in a physically informed manner, where the model determines whether a given galaxy is bright and red enough to be classified as an LRG based on properties related to the size, environment, and formation history of its host halo.

In this final chapter, we use this physically motivated sample to test HOD models, which are one of the standard methods for populating halo catalogues with galaxies within the DESI collaboration. HOD models are calibrated to reproduce the statistical properties of a given sample. However, their approach to determining whether a halo should host an LRG is less physically motivated compared to our NN-based method.

We are particularly interested in assessing how well the galaxy-halo connection of the original sample is preserved through HOD modelling. More specifically, we aim to test whether a set of relationships between bias parameters, introduced in equation~\ref{eq:bconsis} below, remain consistent. These relationships are routinely assumed in RSD analyses within the DESI collaboration \citep[e.g.,][]{2025JCAP...01..134M,2025JCAP...01..129R}.

\subsection{HOD formalism}
\label{HOD_formalism}

We stated that an alternative methodology for painting LRGs into a halo catalogue is to use the HOD formalism. Here, we follow the HOD prescription of \cite{Zheng_2005}, where the probability of a given halo of mass $M_\mathrm{vir}$ containing a central galaxy is given by an error function:

\begin{equation}
\label{central_HOD}
\langle N(M_\mathrm{vir})\rangle=\frac{1}{2} \left(1+\mathrm{erf}\left( \frac{\log_{10}(M_\mathrm{vir})-\log_{10}(M_\mathrm{min})}{\sigma} \right) \right ). 
\end{equation}

Note that $\langle N(M_\mathrm{vir})\rangle$ takes values between zero and one, as each halo can only host one central galaxy. Here, $M_\mathrm{min}$  determines the halo mass at which haloes are 50~per~cent likely to have a central LRG, while $\sigma$ determines the width of the cut-off profile.

Haloes can host more than one satellite LRG. The total number of satellite LRGs is computed using a power-law probability function, which has different slopes for the high and low mass ends.

\begin{equation}
\label{satellite_HOD}
\langle N^{\mathrm{sat}}(M_\mathrm{vir})\rangle=\left( \frac{M_\mathrm{vir}-M_0}{M_1} \right ) ^\alpha \langle N(M_\mathrm{vir})\rangle
\end{equation}

Here $\alpha$ is the slope at the high mass end and $M_0$ and $M_{1}$ regulate the scales where the transition between low and high mass happens. The number of satellites $\langle N(M_\mathrm{vir})\rangle$ is then distributed inside the halo using a NFW profile \citep{1995MNRAS.275...56N}.

To populate a halo catalogue using an HOD model, we determine values for the five free parameters described above, namely $M_\mathrm{min}$, $\sigma$, $M_{1}$, $M_{0}$, and $\alpha$, In what follows, we attempt to find the parameter values that best reproduce the small-scale clustering statistics of our MDPL2 LRG sample.

\subsection{Fitting HODs}
\label{Fitting_HODs}
Our goal is to determine the best-fitting HOD parameters that generate a halo catalogue producing a galaxy sample with a summary clustering statistic that most closely matches those of our LRG MDPL2 mock introduced in Section~\ref{sec:sec_MDPL2_lrgs}.

There are different summary statistics that can be used for this comparison; however, it is common practice to use the projected two-point correlation function $w_{p}$, defined as:

\begin{equation}
\label{wp_definition}
    w_{p}(r_{p})=2 \int ^{r_{\pi,\mathrm{max}}} _0 \xi(r_{p}, r_{\pi}) d r_{\pi},
\end{equation}

where $\xi(r_{p},r_{\pi})$ is the 2-point correlation function, $r_{p}$ and $r_{\pi}$ are the transverse and line-of-sight (LoS) separations, respectively. Throughout this work, $r_{\pi,\mathrm{max}}$ is set to 40\,Mpc\,$h^{-1}$. Each $w_p$ is computed in 12 logarithmically spaced bins between $r_p=0.3$ Mpc\,$h^{-1}$ and $r_{p}=30$\,Mpc\,$h^{-1}$.

$\xi(r_{p}, r_{\pi})$ can be computed from a galaxy sample directly. In this work, this is done using the \cite{1993ApJ...412...64L} estimator:

\begin{equation}
\xi(r_{p},r_{\pi})=\frac{DD(r_{p},r_{\pi})-2DR(r_{p},r_{\pi})+RR(r_{p},r_{\pi})}{RR(r_{p},r_{\pi})(r_{p},r_{\pi})},
\end{equation}

where $DD(r_{p},r_{\pi})$, $RR(r_{p},r_{\pi})$, and $RD(r_{p},r_{\pi})$ are the normalized number of galaxy pairs separated by a transverse distance $r_{p}$ and a LoS distance $r_{\pi}$, corresponding to the data-data, random-random, and data-random pairs, respectively.

Note that $w_{p}(r_{p})$ compresses the information of $\xi(r_{p},r_{\pi})$, thus it losses all information on the velocity distribution encoded along the LoS clustering. However, given that in this form the rest of the clustering information is encoded into a 1-D function and hence any  covariance matrix is significantly simpler to compute, $w_{p}$ is much easier to handle compared to the full 2D correlation function.

In order to compute the $w_p(r_p)$ of an HOD model with a given set of parameters, we use equations~\eqref{central_HOD} and \eqref{satellite_HOD} to populate the full MDPL2 halo sample introduced in Section~\ref{sec:MDPL2_sample} with galaxies. The resulting sample $w_p(r_p)$ can then be compared to the one from our LRG MDPL2 mock. This is done by assuming a Gaussian likelihood and computing a standard $\chi^{2}$ statistic.

\begin{equation}
\label{chi2}
\chi^{2}_{w_{p}}=(w_{p}^\mathrm{HOD}-w_{p}^\mathrm{MDPL2})^{T} C_{w_{p}}^{-1}(w_{p}^\mathrm{HOD}-w_{p}^\mathrm{MDPL2}).
\end{equation}

Here, $w_{p}^\mathrm{HOD}$ and $w_{p}^\mathrm{MDPL2}$ are the projected correlation functions of the resulting HOD sample and of our LRG MDPL2 mock, respectively. $C_{w_{p}}$ is the covariance matrix of our MDPL2 sample, computed using a jackknife method with 125 distinct regions. We construct a set of random points that is 30 times larger than our LRG sample for this purpose.

As stated, we are interested in finding the values of our HOD parameters for which $w_{p}^\mathrm{HOD}$ best fits $w_{p}^\mathrm{MDPL2}$. This is typically done by exploring the HOD parameter space, either using a minimization method or running an MCMC. However, both of these methods require a large number of evaluations of the HOD model at different points in the parameter space. Evaluating a single point can be slow, as one must populate all MDPL2 haloes with galaxies and compute $\xi(r_{p},r_{\pi})$. Both steps can be time-consuming, especially for HOD parameter values that correspond to a high number of satellite galaxies per halo. This makes a thorough exploration of the parameter space prohibitively expensive. A solution that is common in the literature is to build surrogate models \citep[e.g.][]{2020MNRAS.493.5551Y}. Here, we opted to use the NN technology from Section~\ref{sec:NN_train} to build our own surrogate model of $w_{p}$ for different values of the HOD parameter space.

\subsection{Neural Network emulator of \texorpdfstring{$w_{p}^\mathrm{HOD}$}{wp}}
\label{HOD-NN}

The surrogate model is created using the same code as in Section~\ref{sec:NN_train}, but we slightly modify the architecture to better model $w_{p}$. The main change is that we use a total of four hidden layers (instead of two). We also increase the tolerance of the NN to 1000, so training continues until 1000 epochs have occurred without improvement before stopping. Finally, we add an extra step to our methodology: once the emulator reaches its tolerance, we reduce the learning rate by a factor of ten and run the minimization one more time. These changes make our approach very similar to those in other studies that predict two-point clustering statistics of galaxies \citep[e.g.][]{2022JCAP...04..056D,2024JCAP...08..049R}. The NN is trained to reproduce $\log_{10}(rw_{p}(r))$ instead of $w_{p}(r)$ directly. Using the logarithmic function smooths the data and reduces variability, making it better behaved, which simplifies the task for the NN and improves accuracy.

The dataset used to build the model consists of 50,000 measurements of $w_{p}$ across different points in the parameter space. We split this dataset into a training set with 40,000 points, and validation and test sets with 5,000 points each. The parameter space points are selected within the following flat priors: $12.3<\log_{10}(M_\mathrm{min}/\mathrm{M}_{\odot})<14$, $0.01<\sigma<2$, $0.01<\alpha<2$, $12<\log_{10}(M_{0}/\mathrm{M}_{\odot})<14$, and $12<\log_{10}(M_{1}/\mathrm{M}_{\odot})<15$.

The sets of points in parameter space are generated using three distinct Korobov sequences \citep{korobov1959approximate}, designed to sample the parameter space extensively and uniformly.

The Neural Network accurately predicts $\log_{10}(rw_p(r))$ with better than percent-level precision for the points in our test set. However, when the predictions are transformed back from logarithmic space to $w_p(r)$, the errors increase to approximately 1~per~cent. To maintain focus in the main text, we defer the discussion and  accuracy tests of our $w_p(r)$ NN emulator methodology to Appendix~\ref{sec:appendixNN_Hod}.

\subsection{The HOD best fit model}
\label{HOD_bestfit}

Once our NN is built, we can evaluate predictions of $w^{\rm HOD}_p$ in negligible time, which lets us run minimization algorithms to find the parameter set that best reproduces the projected correlation function of our MDPL2 mock. The minimizations are done using the Powell algorithm \citep{2002nrca.book.....P}, which searches for the point in parameter space where $\chi^{2}_{w{p}}$ is minimized. To increase the robustness of the results, we run the algorithm ten times, each time initializing the chain at a different starting point randomly selected within the prior ranges. By running the algorithm multiple times, we reduce the risk of getting stuck in suboptimal local minima. While some minimizations might converge to suboptimal solutions, the others are expected to be close to the global minimum. The final fit is selected as the solution with the lowest overall $\chi^{2}_{w{p}}$ among the ten runs. The validity of this approach is tested further in Appendix~\ref{sec:appendixNN_Hod} bellow.

On small scales, non-central subhaloes significantly affect the clustering of galaxies. While the MDPL2 simulation has sufficient resolution to identify individual subhaloes using \textsc{Rockstar}, HOD models assign subhaloes and distribute them according to an NFW profile. This introduces discrepancies in small-scale clustering, which are challenging to model accurately in the HOD sample. Since small-scale clustering typically has smaller errors, it carries more weight in the fitting process compared to larger scales. Consequently, it is crucial to avoid fitting very small scales where the differences between NFW-distributed and \textsc{Rockstar} subhalo galaxies become more pronounced. Failing to do so may result in models where the larger scales (which have larger errors) are poorly fit as the minimizer prioritizes the smaller scales. We have tested various minimum $r$ scales and found that $r \approx 0.5$ Mpc\,$h^{-1}$ leads to models where the HOD methodology reproduces clustering down to the scales of interest while maintaining accuracy at larger scales (see Section~\ref{PT_work} below).

Fig.~\ref{fig:wp_comparison} shows both the best-fitting emulator prediction of $w^{\rm HOD}_{p}(r)$ (blue dashed line) and $w^{\rm MDPL2}_{p}(r)$ (black dots). The errors in both panels are derived from the diagonal of the jackknife covariance matrix. The bottom panel of the plot demonstrates that the model fits the data well, with fractional errors smaller than those from the diagonal of the covariance matrix. The fit is achieved with the following set of HOD parameters: $\log_{10} (M_\mathrm{min}/\mathrm{M}_{\odot})=12.79$, $\sigma=0.01$, $\alpha=1.24$, $\log_{10}(M_{0}/\mathrm{M}_{\odot})=12.07$, and $\log_{10}(M_{1}/\mathrm{M}_{\odot})=14.03$. The plot also includes the projected correlation function $w^{\rm HOD}_{p}(r)$ measured directly from the HOD methodology (blue solid line), as opposed to using our HOD emulator. The slight difference between the emulator prediction and the actual HOD measurement results in a marginally worse fit to $w^{\rm MDPL2}_{p}(r)$. However, within the jackknife errors, both fits are indistinguishable.

\begin{figure}
\centering
\includegraphics[width=\linewidth]{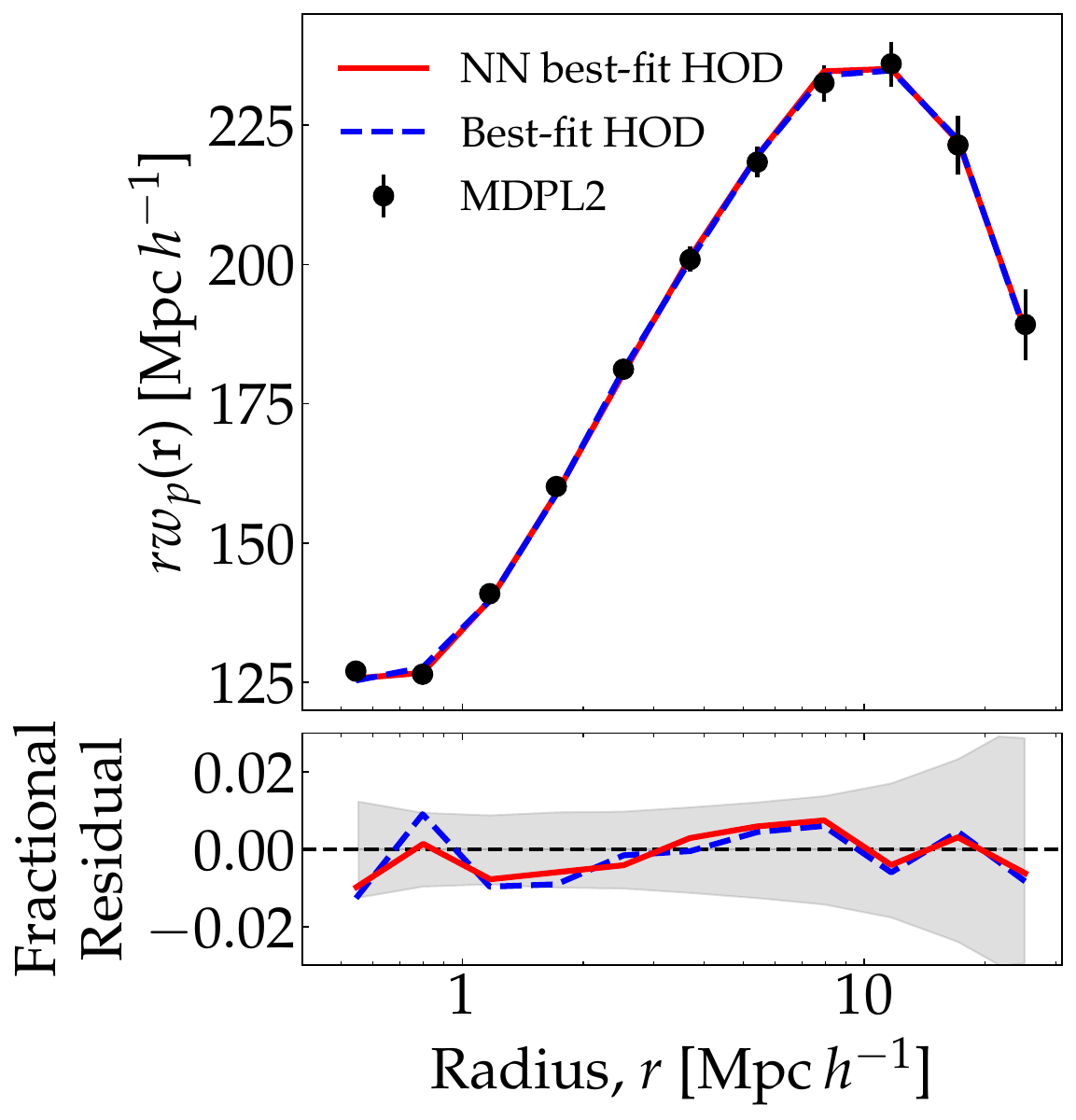}
\caption{Projected correlation function of our best-fitting HOD mock taken from both the HOD emulator (blue dashed line) and measured directly from the HOD sample (red solid line) compared to $w^\mathrm{MDPL2}_{p}(r)$ (black filled circles). The error estimates are approximated as the square root of the diagonal elements of the jackknife covariance matrix for each bin.}
\label{fig:wp_comparison}
\end{figure}

\begin{figure}
\centering
\includegraphics[width=\linewidth]{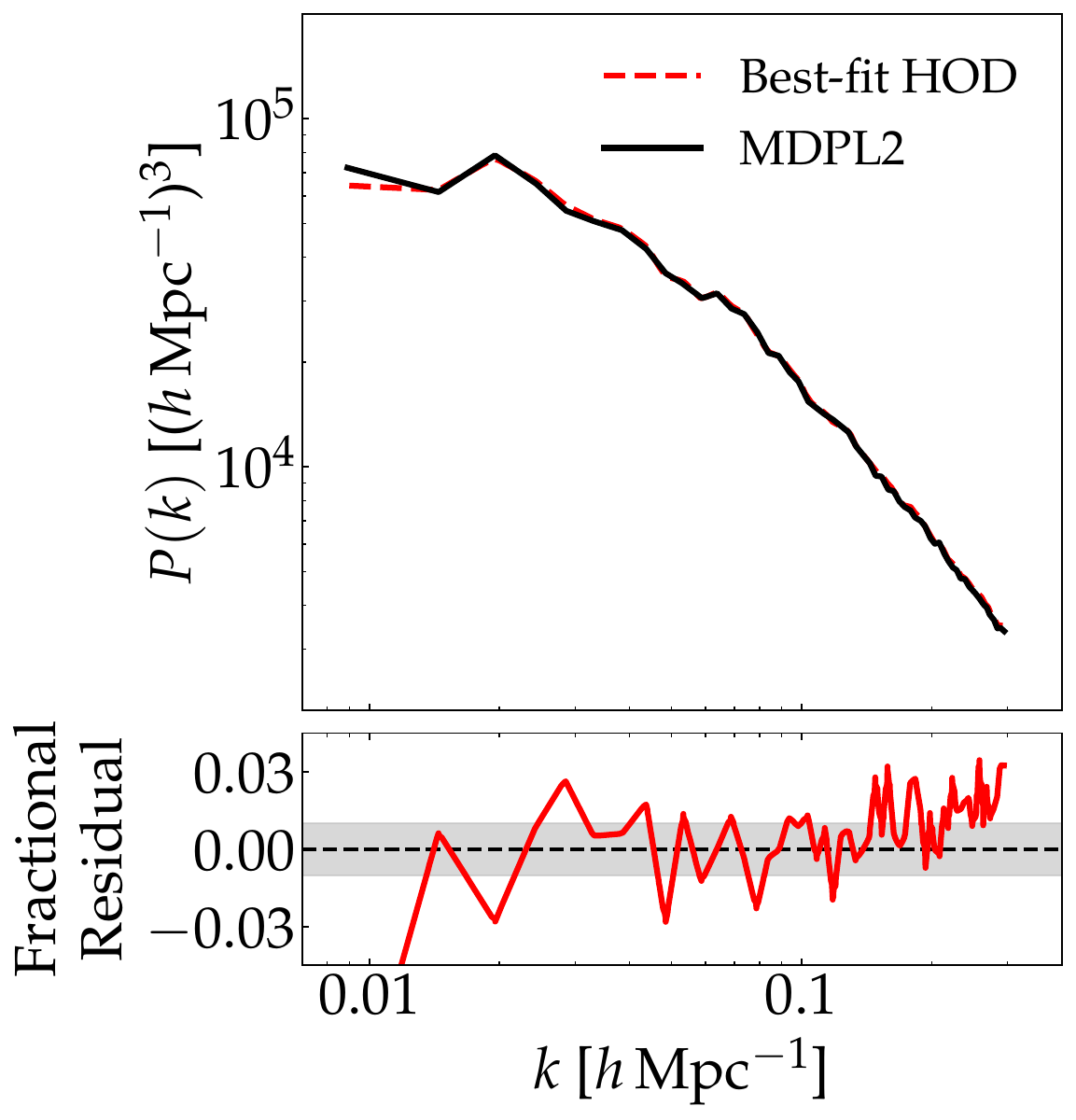}
\caption{Real space power spectra of our MDPL2 LRG sample (black solid line), and the best-fitting HOD model (red dashed line). The bottom panel shows the fractional residual of the model and sample.}
\label{fig:Pk_comparison}
\end{figure}

Fig.~\ref{fig:Pk_comparison} shows the real-space projected correlation function for both our MDPL2 LRG sample and our best-fitting HOD sample. We find that the two curves agree well over the range $k \sim 0.02\,h$\,Mpc$^{-1}$ to $k \sim 0.17\,h$\,Mpc$^{-1}$, indicating that the clustering remains consistent even on large scales. It is important to note that, by construction, the HOD model was fitted to reproduce the small-scale clustering of galaxies but not necessarily the large-scale clustering. The fact that the agreement extends to large scales demonstrates the robustness of HOD models, even when using relatively simple prescriptions like ours, that depend only on stellar mass.

\subsection{Consistency Tests of Perturbation Theory Bias Relations}
\label{PT_work}

\begin{figure*}
\centering
\includegraphics[width=86mm]{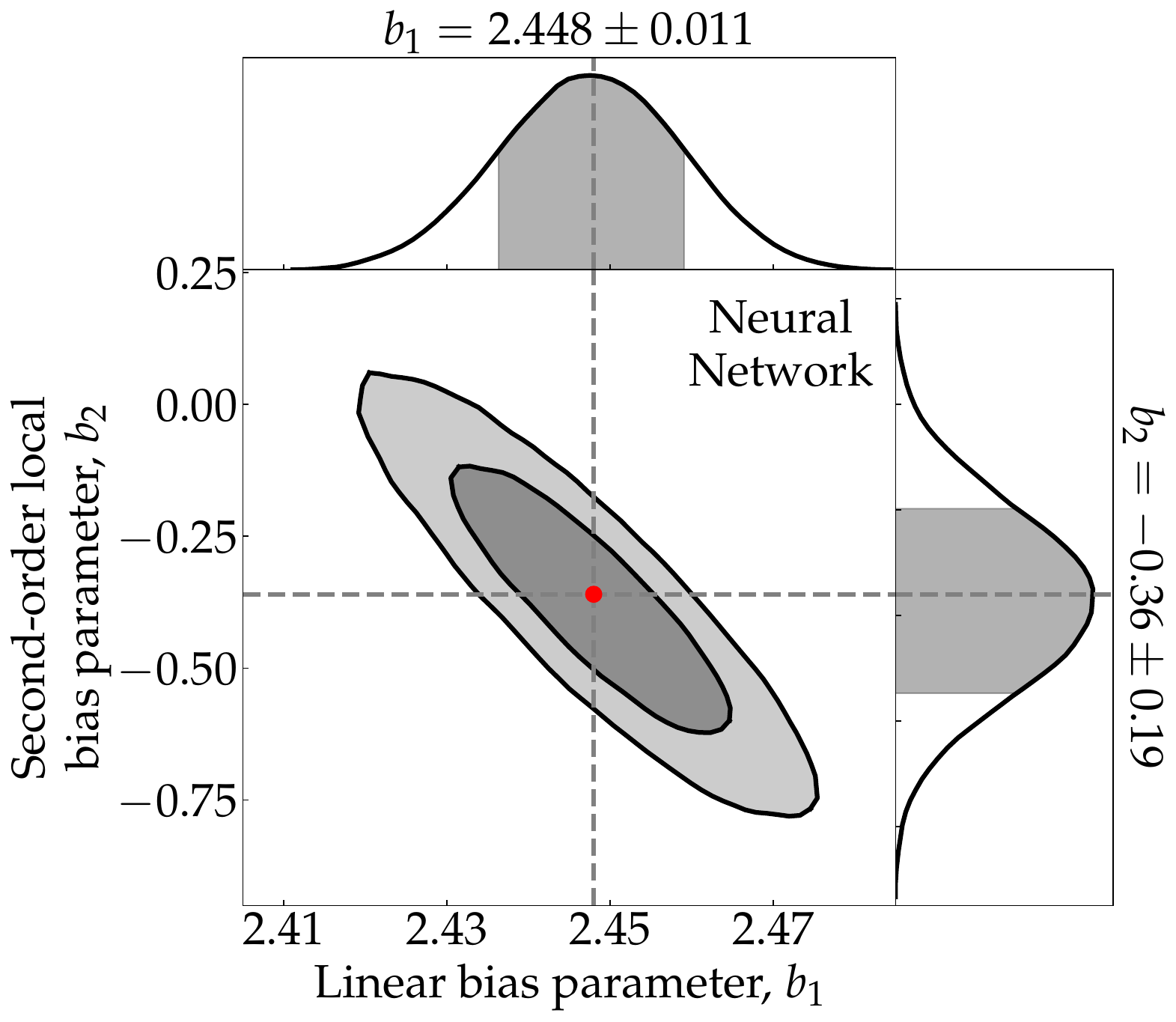}\hfill
\includegraphics[width=86mm]{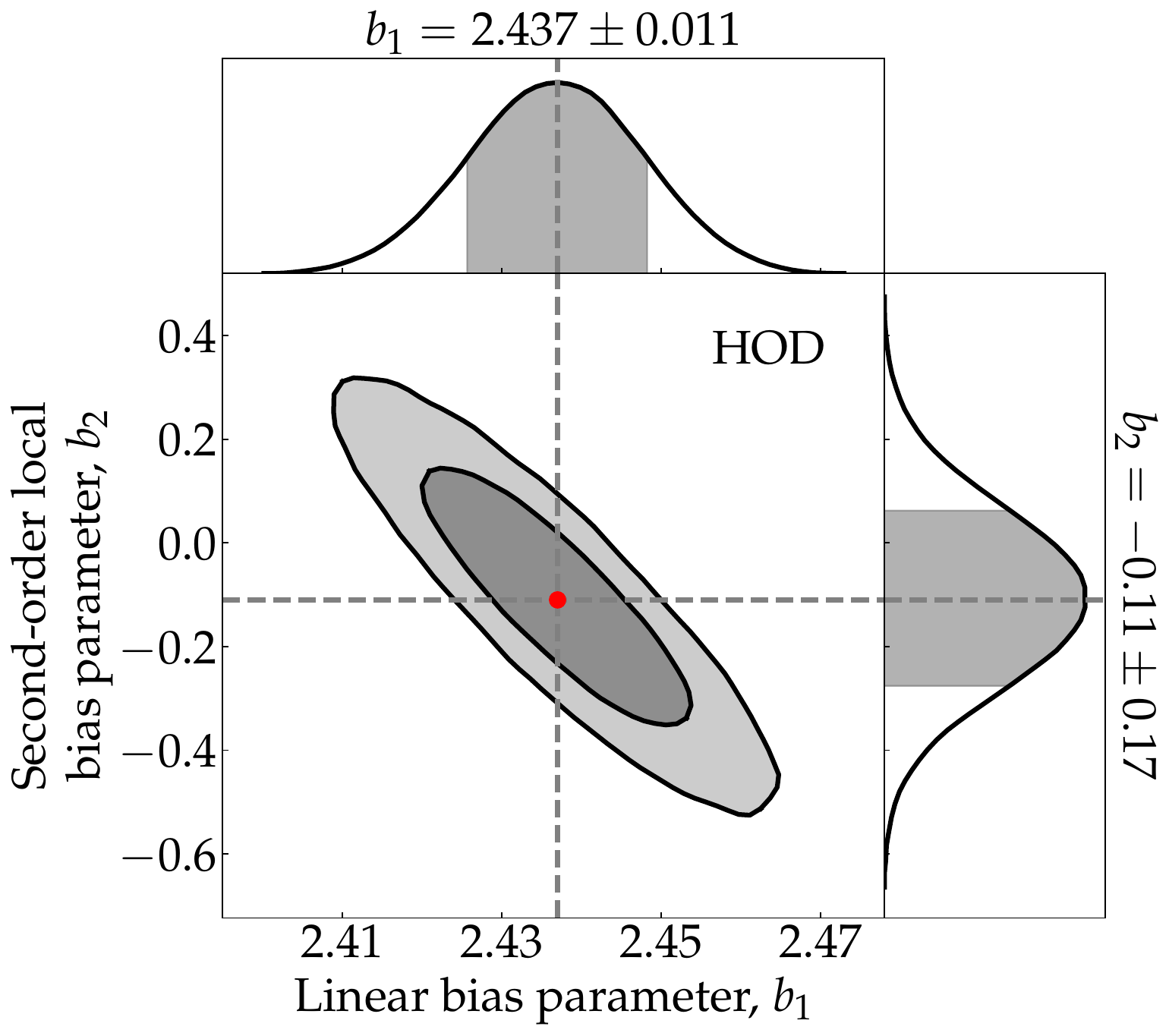}
\caption{2D joint posterior distribution of parameters $b_{1}$ and $b_{2}$, obtained from MCMC sampling of our bias model applied to the MDPL2 NN sample (left) and the HOD sample (right), assuming that the higher-order parameters are determined by the consistency relations, see equation~\ref{eq:bconsis}. The histograms along the top and right show the marginalized 1D posterior distributions for each parameter. The shaded regions under the histograms represent the 68~per~cent highest posterior density intervals.}
\label{fig:b1_b2}
\end{figure*}

We have generated two different sets of LRG samples; the MDPL2 LRG sample, constructed in Section~\ref{sec:data_MDPL2} using our NN colour and magnitude predictions, and a second sample derived by fitting an HOD model to the former. We have shown that while we only require the small-scale clustering to be consistent between the two samples, the large-scale clustering is also consistent. This serves as a good sanity check of our methodology, though it is a well-known property of an efficient HOD parametrization.

Our methodology places us in a unique position, as we can test how well the HOD model recovers the galaxy-halo connection of the original sample. This is typically challenging because observational surveys cannot directly observe the host halos of galaxies. In clustering analyses, the galaxy-halo connection is commonly incorporated into the modelling using approaches such as second-order perturbation theory. This introduces additional nuisance parameters that must be fitted alongside cosmological parameters. Previous HOD analyses have shown that, when fitting these models to an HOD-generated sample, some bias parameters exhibit strong correlations with one another \citep[e.g.][]{2024PhRvD.110f3538I,2024arXiv240912937Z}. These correlations imply that certain bias parameters can be expressed as functions of others. This allows for a reduction in the number of free parameters explored in clustering analyses, simplifying the process and mitigating prior effects.

However, we note that HOD models are generally simpler and have fewer free parameters than the perturbation theory models commonly used in clustering studies. As a result, they may introduce artificial parameter correlations that do not necessarily appear in more complex, physically motivated models, such as the one used to generate our NN sample. In this work, we are uniquely positioned to test whether the original sample used to fit our HOD also adheres to these relationships.
 
We now introduce a second-order perturbation bias model, where the relationship between DM particles and galaxies is described using a non-linear and non-local galaxy bias framework. Here, we adopt the model from~\cite{2009JCAP...08..020M}, which has been applied to the data analyses of several galaxy survey experiments such as Euclid \citep{2024A&A...689A.275E}. The density of galaxies is expressed as a function of the DM density, $\delta_g(\delta)$. Thus, the density field of the galaxy distribution, $\delta_g(\bfx)$, is expanded in terms of the DM density field, $\delta(\bfx)$, and its tidal tensor field, $s(\bfx)$, as:
\bea
\label{eq:bias_x}
\delta_g(\bfx)&=&b_1\delta(\bfx)+\frac{1}{2}b_2[\delta(\bfx)^2-\sigma_2]+\frac{1}{2}b_{s2}[s(\bfx)^2-\langle s^2\rangle]\nonumber\\
&+&{\rm higher\,\,order\,\, terms}.
\eea
The tidal tensor fields are expressed as $s(\bfx)=s_{ij}(\bfx)s_{ij}(\bfx)$, with $s_{ij}(\bfx)=\partial_i\partial_j\Phi(\bfx)-\delta_{ij}^{\rm Kr}\delta(\bfx)$, and where $\Phi(\bfx)$ represents the gravitational potential. Also, the linear bias parameter, the second--order local bias parameter and the second-order non-local bias parameter are denoted by $b_{1}$, $b_{2}$ and $b_{s2}$ respectively. 

The galaxy density field $\delta_g(\bfx)$ in equation~\ref{eq:bias_x} is transformed into Fourier space as,
\bea
\label{eq:bias_k}
\delta_g(\bfk)&=&b_1\delta(\bfk)+\frac{1}{2}b_2\int \frac{d\bfq}{(2\pi)^3}\delta(\bfq)\delta(\bfk-\bfq)\nonumber\\
&+&\frac{1}{2}b_{s2}\int \frac{d\bfq}{(2\pi)^3}\delta(\bfq)\delta(\bfk-\bfq)S_2(\bfq,\bfk-\bfq)\nonumber\\
&+&{\rm higher\,\, order\,\, terms},
\eea
where the tidal kernel $S_2(\bfq,\bfk-\bfq)$ is defined as,
\beq
S_2(\bfk_1,\bfk_2) \equiv \frac{(\bfk_1\cdot\bfk_2)^2}{(k_1k_2)^2}-\frac{1}{3}.
\eeq

\noindent The density field power spectrum $P_{\delta_g\delta_g}$ is given by
\bea
P_{\delta_g\delta_g}&=&P^2_{\delta_g\delta}/P_{\delta\delta},
\label{eq:pdd_bias}
\eea
where the cross--spectrum between galaxy and DM is expressed up to one--loop order and is written as
\bea
\label{eq:phm_bias}
P_{\delta_g\delta}(k)&=&b_1P_{\delta\delta}(k)+b_2P_{b2,\delta}(k)+b_{s2}P_{bs2,\delta}(k)
\nonumber \\
&+&b_{3\rm{nl}}\sigma_3^2(k)P^{\rm{L}}_{\rm m}(k),
\eea
where the bias potentials $P_{b2,\delta}$ and $P_{bs2,\delta}$ are given at~\cite{2009JCAP...08..020M}, and $P^{\rm{L}}$ is the linear dark matter spectrum. The third--order non--local bias parameter is denoted by $b_{3\rm{nl}}$. The locality of density bias leads to the following consistency relation among bias parameters,
\bea\label{eq:bconsis}
b_{s2}=-\frac{4}{7}(b_1-1) \,,
&\quad&
b_{3\rm{nl}}=\frac{32}{315}(b_1-1)\,.
\eea

In this section, we aim to test the consistency of these relationships for both of our LRG samples. We begin by fitting the theoretical perturbative bias models described above to the two measured galaxy power spectra from Section~\ref{HOD_bestfit} independently. The fits are performed up to $k_\mathrm{max} = 0.17\,h$\,Mpc$^{-1}$, which represents a reasonable scale limit for accurate measurements of cosmological parameters in usual clustering analysis. We fix the cosmological parameters to their fiducial values, while both local and non-local bias parameters are varied during the fit.

With this relatively low $k_{\rm max}$ bound, the non-local bias parameters are not measured very precisely, resulting in large error bars in the estimation of the parameters. Despite these weak constraints, we find that the relation from equation~\ref{eq:bconsis} holds. However, as mentioned, the constraints are broad, so this does not constitute strong evidence for the relation. For instance, the estimated value by the consistency relation is $b_{s2}=-0.8$, but the measured $b_{s2}=-0.07 \pm 2.0$.

A secondary test to determine whether the HOD model affects the galaxy-halo connection is to assess whether assuming the consistency relations from equation~\ref{eq:bconsis} leads to different results when fitting the HOD-generated LRG sample compared to our physically informed MDPL2 sample. To do this, we perform a new fit where we assume the consistency relations hold for $k_{\rm max} < 0.17\,h$\,Mpc$^{-1}$ in both LRG samples, fixing the non-local parameters to the values given in equation~\ref{eq:bconsis}. The measured local bias parameters of $b_{1}$ and $b_{2}$ are presented in Fig.~\ref{fig:b1_b2} for both LRG samples of MDPL2 and HOD in the left and right panels respectively. The HOD galaxies populated to mock characteristics of projected correlation function of MDPL2 sample at much smaller scales, exhibit the similar bias parameters observed at quasi--linear scales. The leading order local bias parameter $b_{1}$ is nearly identical for both samples, and the first order local bias parameter $b_{2}$ is measured within the $1\sigma$ range as well. 

\section{Conclusions}
\label{conclussions}

The analysis of LSS surveys requires highly accurate mock galaxy catalogues. In large-volume simulations, where galaxy clustering remains well-defined even at scales of hundreds of Mpc, hydrodynamical simulations become prohibitively expensive. As a result, these simulations are typically DM-only, and galaxies must be assigned to haloes a posteriori. A common approach for this task is the use of HOD models, which statistically reproduce key properties such as small-scale clustering. In this work, we explore an alternative approach using NNs to model the complex, physically motivated relationships between host haloes and their galaxies.

The NNs are trained on the largest IllustrisTNG hydrodynamical simulation, which has a side length of approximately 300\,Mpc. The network predicts three galaxy properties, stellar mass ($\mstar$), the $z^{\prime}$ band magnitude, and the ($r^{\prime}-z^{\prime}$) colour, based on nine dark matter (DM) parameters. These parameters include a flag indicating whether the subhalo is central or satellite, V$_\mathrm{max}$, M$_\mathrm{max}$, M$_{z=0.82}$, the four FC$_{[30, 50, 70, 90]}$ parameters, and the redshift at which the subhalo reaches its maximum mass.

All galaxy properties are extracted from the hydrodynamical simulation, while the DM parameters are taken from a corresponding DM-only simulation built with the same initial conditions but without baryons. A pre-existing halo matching between the hydrodynamical and DM-only simulations ensures that any baryonic effects present in the former do not influence the haloes in the latter, maintaining consistency with the large-volume DM-only simulations.

To evaluate the performance of our NN predictions, we use them to estimate the properties of galaxies in the test set. At first glance, the predictions appear relatively accurate, as the Pearson correlation coefficients between the predicted and true values are fairly high, with ($r^{\prime}-z^{\prime}$) and $z^{\prime}$ yielding $\rho_{xy}\sim0.8$ and $\mstar$ reaching $\rho_{xy}\sim0.96$. However, we observe a slight asymmetry in the distribution of predictions. Our model tends to slightly underestimate the brightness of the brightest galaxies, overestimate the redness of the reddest galaxies, and systematically under-predict $\mstar$ more often than it over-predicts it.

As a consequence of these asymmetries, we cannot apply the same selection cuts for LRGs as those used in our {\it true} sample, as this would alter the total LRG density, which we aim to match with the DESI sample. To address this, we use the validation set to make a slight adjustment to the selection cuts, ensuring that the final LRG density matches the desired value while also maximizing the fraction of {\it true} LRGs selected.

We aim to quantify how similar the TNG300 sample, generated by populating our test set (which contains 20~per~cent of our data) using our NN models, is to the original {\it true} sample. To do this, we compare both samples statistically. Both exhibit very similar LRG densities, which is expected since the model is designed to replicate the true density of LRGs in the validation set. However, we find that the TNG300 sample contains a higher fraction of massive haloes as LRG hosts.

We believe this discrepancy arises because we are populating a DM-only simulation, which lacks information from baryonic processes such as AGN feedback or other mechanisms that can quench galaxies in smaller haloes. As a result, our NN model tends to predict large and old subhaloes as red and bright, while smaller subhaloes that may have been reddened through baryonic processes are not included.

We use this model to populate the large MDPL2 simulation with a side length of 1\,Gpc\,$h^{-1}$ with LRG galaxies. The resulting galaxy sample is statistically similar to the TNG300 sample. The main difference we find is that the MDPL2 sample contains a larger fraction of LRGs hosted by central subhaloes. The origin of this discrepancy is unclear to us, but we suspect it may be generated, at least in part, from intrinsic differences between the simulations. In particular, we observe that MDPL2 generally contains a higher fraction of central subhaloes than the TNG300 sample in the relevant mass range, even before LRGs are assigned.

We aim to use our large-volume MDPL2 LRG sample to study how well HOD models preserve the galaxy-halo connection. This is a unique opportunity, as HOD models are typically fitted to observational LRG samples, where the host halo information is not directly available. These models provide a statistical prescription for the galaxy-halo connection without direct validation against true halo properties. However, since we have full knowledge of the host haloes in our MDPL2 LRG sample, we can run an HOD analysis on this dataset and directly verify whether the galaxy-halo connection remains consistent between the HOD-generated and original samples.

We use a relatively simple HOD prescription that depends only on the mass of the host halo and assigns satellite LRGs following an NFW profile. While a more complex model could be used, we demonstrate that even with this straightforward approach, the galaxy-halo connection is preserved well enough. We fit the HOD parameters to reproduce the small-scale projected correlation function, $w_{p}$, of the MDPL2 sample.

Predicting $w_{p}$ for a new point in parameter space is computationally expensive, as the code must first populate all individual subhaloes with galaxies and then compute the correlation function for the sample. To improve efficiency, we built a NN emulator that acts as a surrogate model for $w_{p}$. This significantly reduces the number of direct evaluations required, bringing it down to 50,000 evaluations needed to generate the training, test, and validation sets for the NN model, which can reproduce $w_{p}$ with percent-level accuracy.

As stated, we aim to test whether the galaxy sample built using the best-fitting HOD models accurately preserves the galaxy-halo connection. More specifically, we examine whether the correlations between the bias parameters $b_{1}$, $b_{s2}$ and
$b_{3\rm{nl}}$ as given by the consistency relations introduced in the main text, remain consistent between the HOD-generated and original samples. This test is important because these correlations suggest that $b_{s2}$ and $b_{3\rm{nl}}$ are not independent but instead fully determined by the value of $b_{1}$. This assumption is routinely used in DESI LRG analyses to reduce the number of free parameters and simplify the parameter space exploration.

We begin by computing the power spectrum for both samples independently. We then determine the best-fitting values of the four free parameters, $b_{1}$, $b_{2}$ $b_{s2}$ and $b_{3\rm{nl}}$, that yield power spectrum models best matching the measured spectra for each sample. We find that the retrieved parameter values follow the expected correlations. However, the uncertainties are large, making it difficult to draw definitive conclusions.

As a secondary test, we measure whether assuming these relations affects our results when using HOD models. To do this, we refit our perturbation theory model to both samples, but this time, we vary only $b_{1}$ and $b_{2}$ while fixing $b_{s2}$ and $b_{3\rm{nl}}$ to the values given by the consistency relations. We find almost no differences in the estimated values and uncertainties of $b_{1}$, while the predicted values of $b_{2}$ agree to within 1$\sigma$.

\section{Data Availability}

The authors are committed to share the data used in this work if requested. The TNG300 data is publicly available\footnote{See \url{https://www.tng-project.org/data/}. Note that an account is required.}. The data from MDPL2 is also publicly available and can be accessed trough the CosmoSim database\footnote{\url{https://www.cosmosim.org/cms/simulations/mdpl2}.} \citep[for more information, see][]{Riebe_2011, Prada_2012}.

\section*{Acknowledgements}
We sincerely thank Dr. Boryana Hadzhiyska and Dr. Sandy Yuan for kindly providing their magnitude catalogues for the Illustris simulation, which were instrumental in this work. Their contributions and support are greatly appreciated. MIL acknowledges the support from Comunidad de Madrid in Spain (grant Atracci\'{o}n de Talento Contract no. 2023-5A/TIC-28943). ELS was supported by the Australian Government through the Australian Research Council Centre of Excellence for Dark Matter Particle Physics (CDM, CE200100008). This work was supported by the high performance computing cluster Seondeok at the Korea Astronomy and Space Science Institute (KASI). F. S. acknowledgements supports from the National Natural Science Foundation of China (Nos.12103037) and the science research grants from the China Manned Space Project (Grant No. CMS-CSST-2021-A02). 

\bibliographystyle{mnras}
\bibliography{references} 

\appendix

\section{Limitations of the data}\label{sec:appendixA}

\subsection{Missing subhaloes}

The data from our {\it magnitude galaxy sample} contains 80,000 galaxies that have a stellar mass between $10 < \log_{10}(\mathrm{M_{\bigstar}/M_{\odot}}) < 12$. However, these are not all the TNG300 galaxies inside this stellar mass range; there are 2,590 galaxies missing. 

\begin{figure}
\centering
\includegraphics[width=\linewidth]{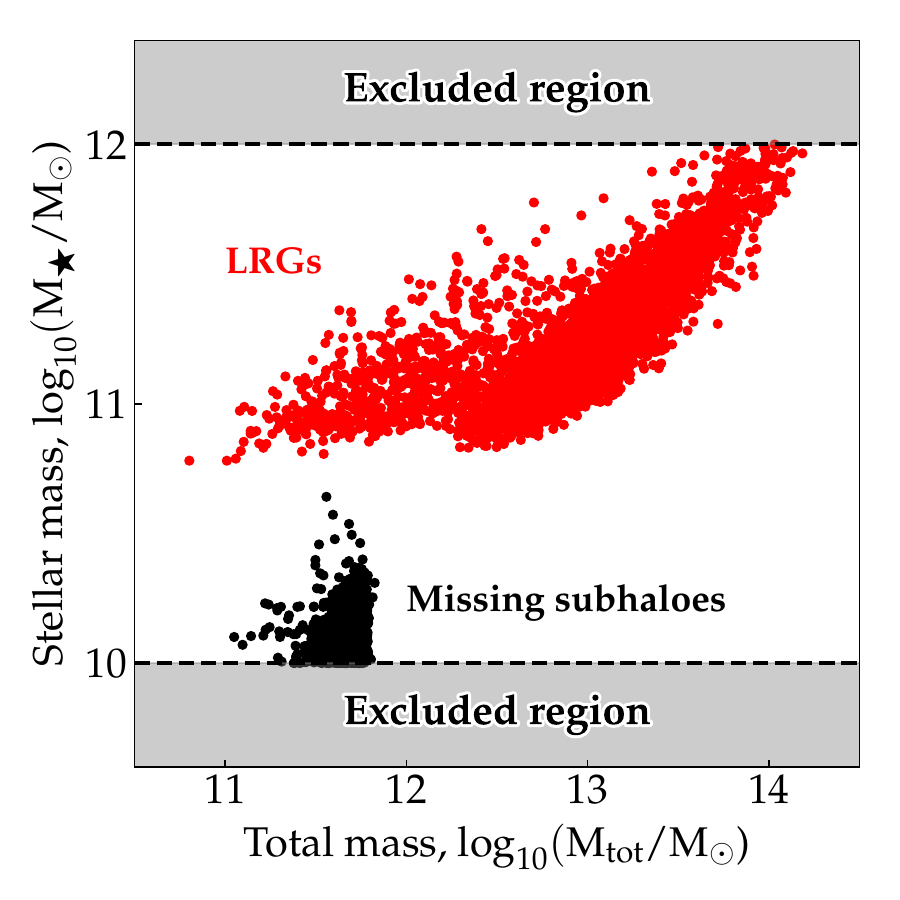}
\caption{Log of stellar mass vs. log of total mass for the LRGs in our sample (red points) and the subhaloes with $10 < \log_{10}(\mathrm{M_{\bigstar}/M_{\odot}}) < 12$ excluded from our sample (black points). Each set of subhaloes occupies a distinct location in the parameter space.}
\label{fig:missing_subs}
\end{figure}

Fig.~\ref{fig:missing_subs} shows the stellar mass and total subhalo host mass of the 2,590 galaxies (black dots) that were excluded from the colour sample generated by \citet{2022MNRAS.512.5793Y} and compares them with the LRG sample from the same set (red dots). The missing galaxies mostly correspond to galaxies living in small subhaloes with $11.5 < \log_{10}(\mathrm{M_{tot}/M_{\odot}}) < 12$. The plot shows that both subsets are independent in parameter space, having a clean separation in stellar mass.

While the lower mass tail of the LRG subhalo mass overlaps with the average subhalo mass of the missing galaxies, the differences in stellar mass between the red and black dots are larger than $\Delta \log_{10}(\mathrm{M_{\bigstar}/M_{\odot}}) \gtrsim 0.5$. We note from Fig.~\ref{fig:pred_mstar}, that the error in our NN predictions of the stellar mass are much smaller than this. Therefore, it is highly unlikely that any of the missing haloes would have been selected as an LRG by our model, and as such including these galaxies would probably have no effect in our resulting LRG statistics. We also note that the number of missing galaxies is small compared to the total number of galaxies in our training set. Hence, the improvements we might achieve from a larger training set are probably minimal.

\subsection{Differences in mass definition}

The subhalo mass is defined differently in TNG300 and MDPL2. In TNG300, the total subhalo mass is the total mass of all particles associated to the subhalo by the \textsc{subfind} algorithm, while in MDPL2 the total mass is the combined mass of all DM particles inside the viral radius of the subhalo.

\begin{figure}
\centering
\includegraphics[width=\linewidth]{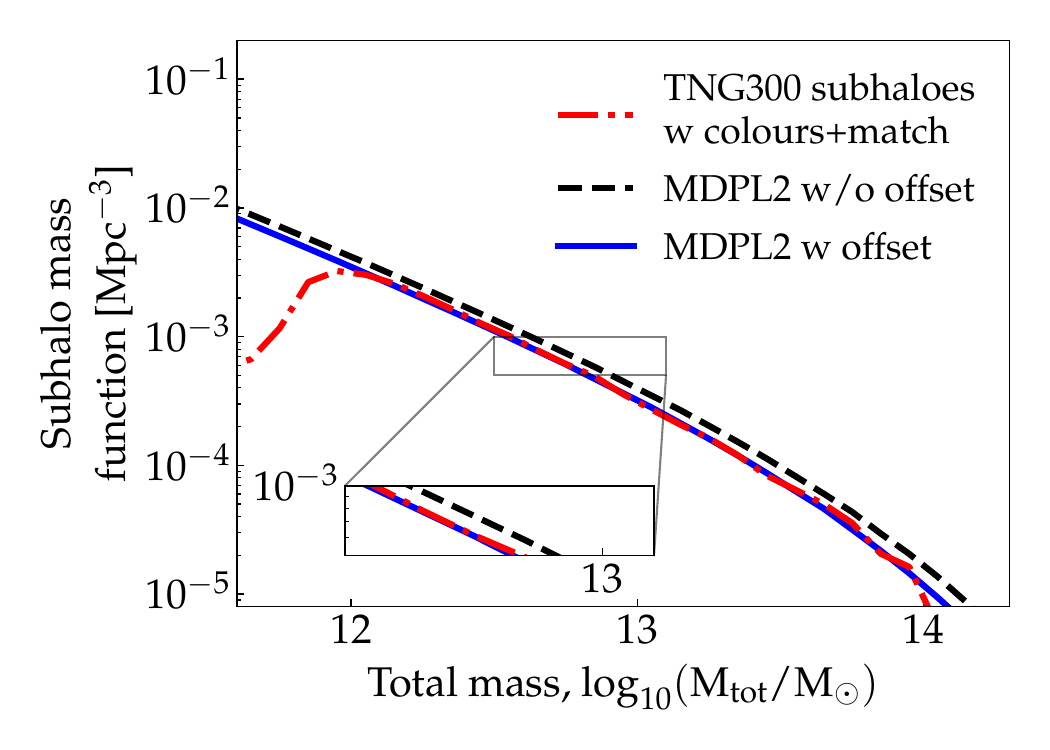}
\caption{The subhalo mass function of TNG300 subhaloes (red dashed-dotted, the same as in Fig.~\ref{fig:completeness}), of MDPL2 subhaloes without any correction to the subhalo mass (black dashed), and of MDPL2 subhaloes with a correction of $\sim$0.082\,dex to the (log) subhalo mass (blue solid).}
\label{fig:SHMF_offset}
\end{figure}

This difference in definition leads to a small offset in the subhalo mass. As a result, the distribution of masses is similar between both simulations but the TNG300 subhaloes tend to be slightly larger than their MDPL2 counterparts. This can be seen in Fig.~\ref{fig:SHMF_offset}, where the sHMF of our TNG300 data (red dashed-dotted line) has a similar shape to the sHMF of MDPL2 (black dashed line), but is systematically below it. 

To ensure that the resulting LRG samples are consistent, we need equivalent parameter distributions between both simulations. With this in mind we find the best-fitting offset to the MDPL2 subhaloes (decreasing their mass) that makes the sHMFs as similar as possible. We use the Powell algorithm to minimize the mean ratio between the TNG300 subhalo functions and the calculated MDPL2 function with the offset applied such that it approaches a ratio of one, i.e.\ when the two distributions are equal. We fit the sHMF in the region where $ 12.3 < \log_{10}(\mathrm{M_{tot}/M_{\odot}}) < 13.5$ to avoid the regions where the TNG300 sHMF drops off. With this method, we find a best-fitting offset of 0.082\,dex. The sHMF for the MDPL2 subhaloes with the offset applied is shown as the blue line of Fig.~\ref{fig:SHMF_offset}.

\section{Impact of Data Partitioning}\label{sec:appendixB}

Our methodology for selecting LRGs has some intrinsic variation due to the randomness involved when dividing our data into training, test, and validation sets. This effect is twofold: it determines both the data that will be left out of the training of the NN and the subset of LRGs in the validation set that will determine our best-fitting cuts.

To test the effect of these variations in our methodology, we ran our entire process five times using the same configuration described in Section~\ref{sec:NN_train}, each time with a different random state for selecting our test and validation samples. 

We find no significant difference between the actual predictions of the NNs with different seeds. We find the MSE to be consistent between the five runs, with stellar mass having a mean MSE of 0.0950 with a variation of 0.0007, for $z^{\prime}$ we find a mean MSE of 0.354 with a variation of 0.003, and finally for ($r^{\prime}-z^{\prime}$) we find a mean MSE of 0.1917 with a variation of 0.0007. I.e.\ the variation between the five runs is below one~per~cent for all prediction parameters.

\begin{figure}
\centering
\includegraphics[width=\linewidth]{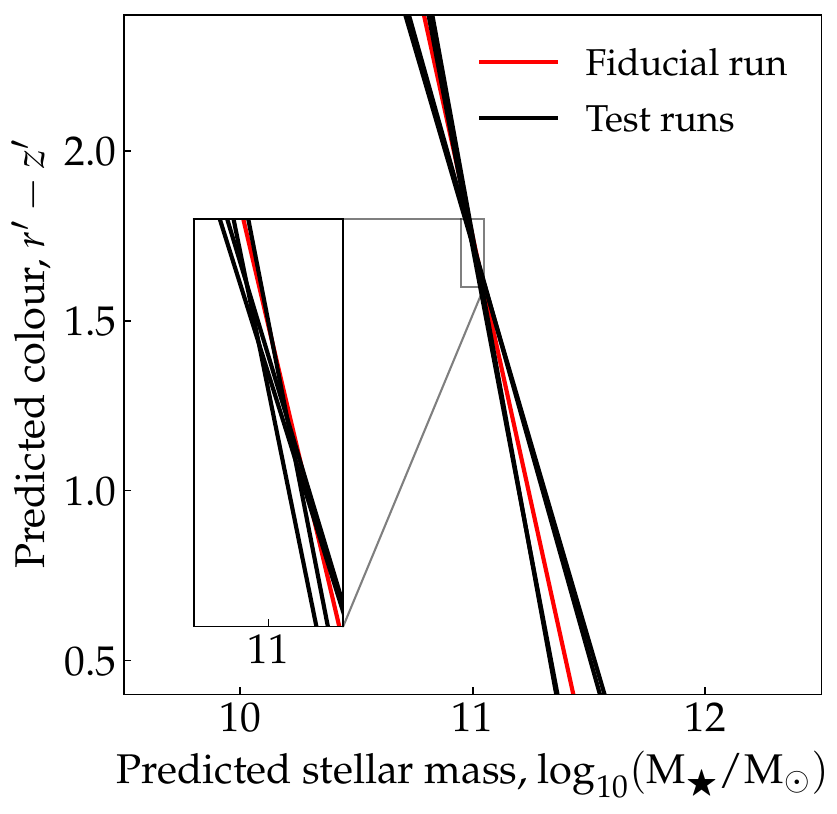}
\caption{The best-fitting cuts to the colour-stellar mass plane for the five runs of our NN (four test runs plus the fiducial run utilised in the paper). The red line shows the best-fitting cut of Fig.~\ref{fig:SM_r_z_cut}, while the black lines show the cuts for the test runs. The inset shows that the five lines do not intersect at one singular point.}
\label{fig:test_cuts}
\end{figure}

We also find low variation in the best-fitting cuts to the (predicted) stellar mass-colour plane of the validation set of each run. Fig.~\ref{fig:test_cuts} shows best-fitting cuts for all five runs, with the red line indicating the cut of the NN used throughout the paper (this line is identical to the black dashed line of Fig.~\ref{fig:SM_r_z_cut}).

The slope and intercept of the cuts show little variation, as shown in the inset of Fig.~\ref{fig:test_cuts}, but are not identical.

\section{Assessing the HOD Surrogate Model and Parameter Estimation}\label{sec:appendixNN_Hod}

\begin{figure}
\centering
\includegraphics[width=\linewidth]{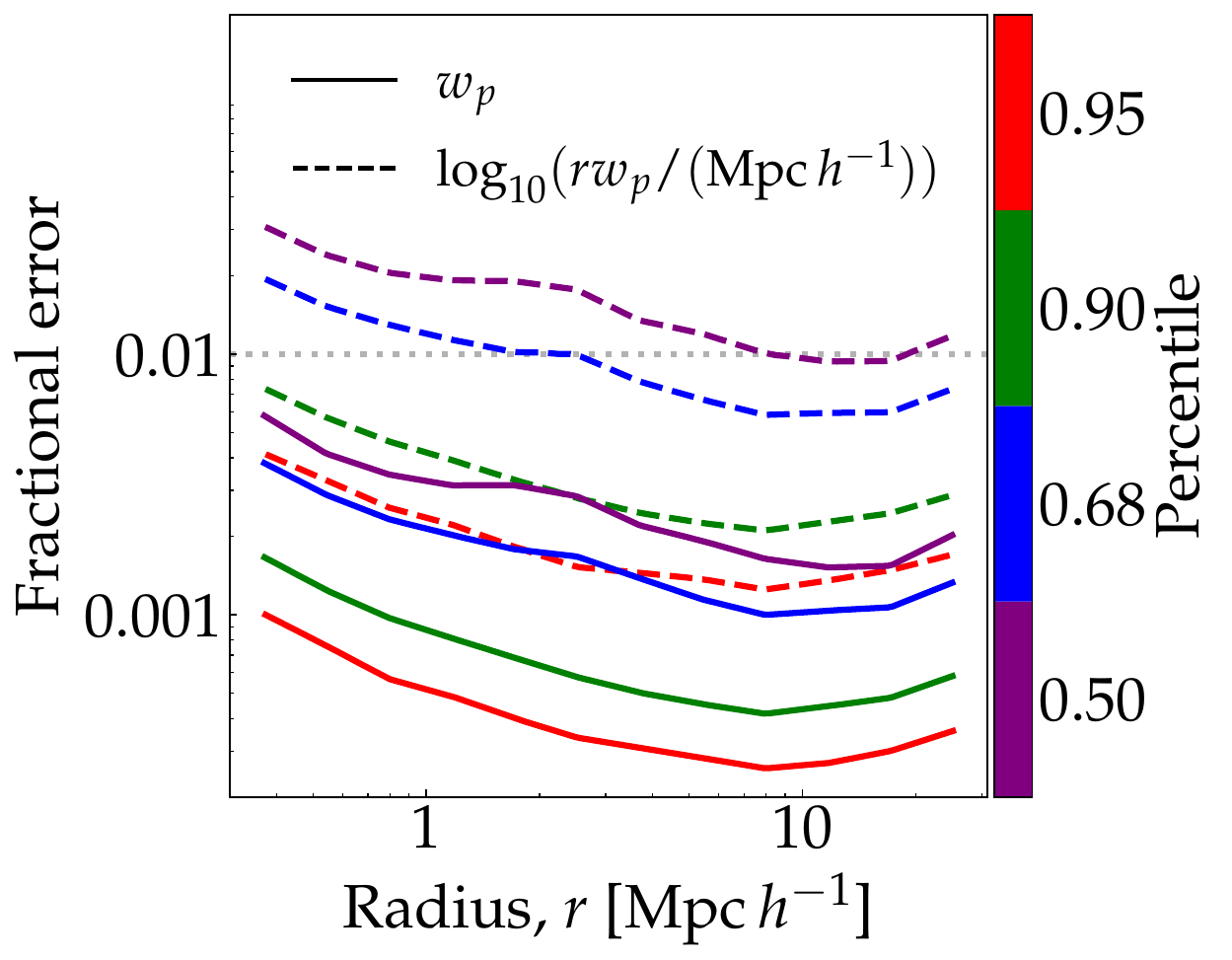}
\caption{The solid coloured lines represent the threshold below which a given percentile of the percentage errors of the emulator predictions fall. The emulator predicts $\log_{10}(rw_{p}(r))$, and the dashed lines show how these same errors transform when used to predict $w_{p}$ instead. The grey dotted line represents the one~per~cent error threshold.}
\label{fig:nn_hod_accracy}
\end{figure}

In Section~\ref{HOD-NN}, we introduced a neural network (NN) emulator for the projected correlation function. In this appendix, we present quantitative tests to evaluate the accuracy of these surrogate models. The solid lines in Fig.~\ref{fig:nn_hod_accracy} illustrate the percentile thresholds below which
50, 68, 90, and 95~per~cent of the per~cent errors for all $\log_{10}(rw_p(r))$ predictions in the test set lie for a given scale $r$. The plot demonstrates that most predictions exhibit very small percent errors, with 50~per~cent of the points having a percent error below 0.1~per~cent, and the remaining predictions showing errors smaller than 1~per~cent. As stated in the main text, the NN is trained on $\log_{10}(rw_{p}(r))$ to ensure smoother data and to simplify the task of the NN.

The dashed coloured lines represent how the percent errors transform when converting the $w_{p}(r)$ predictions back from logarithmic space. As expected, these lines indicate larger percent errors; however, 68~per~cent, of the test set predictions still exhibit errors below 1~per~cent, and 90~per~cent of the points maintain errors below 2~per~cent, even at the smallest scales. 

We are also interested in testing the minimization methodology used to determine the best-fitting HOD parameters that fits the NN predictions of $w_{p}$ to the $w_{p}$ of our MDPL2 mock. To do this, we create a new LRG mock sample by populating the MDPL2 halo catalogue with galaxies using our HOD methodology, setting its parameters to the fit values obtained from the main analysis. We refer to this galaxy sample as the {\it best-fitting} HOD mock.

After constructing the catalogue, we measure its $w_{p}$, which we then attempt to fit from scratch using our minimization methodology to determine the best-fitting HOD parameters for this new mock. Since the HOD parameters used to generate this mock are known, we can evaluate the ability of our methodology to recover the original parameters and assess how accurately the model reproduces $w_{p}$. Note that this requires a new covariance matrix, which we construct using our jackknife method applied to the {\it best-fitting} HOD mock.

\begin{figure}
\centering
\includegraphics[width=\linewidth]{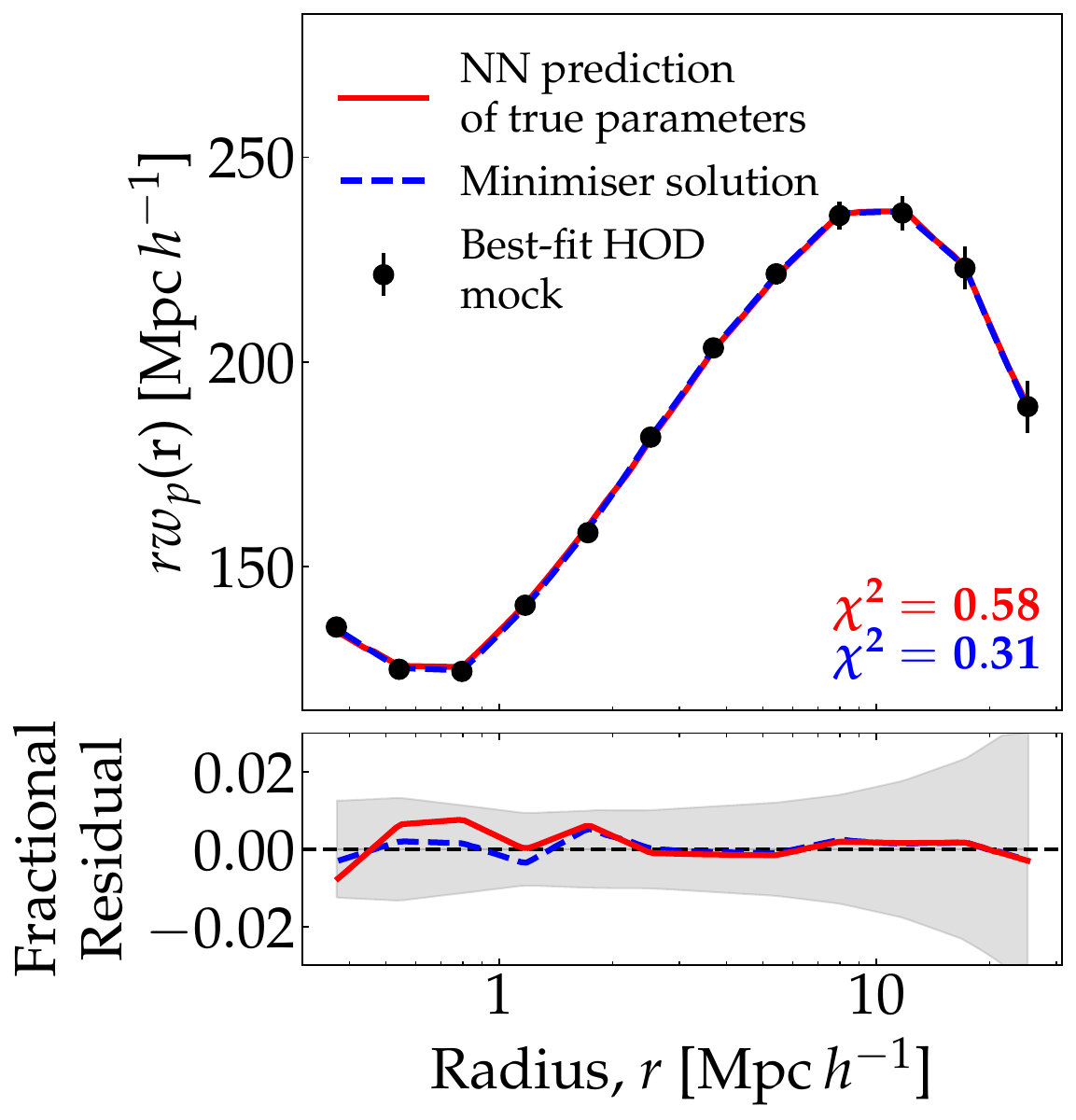}
\caption{Projected correlation function of our {\it best-fitting} HOD mock (black dots) compared to the emulator's prediction of $w_{p}$ using the true parameters used to build the mock (red solid line) and the solution found by our minimization methodology (blue dashed line). The error estimates are approximated as the square root of the diagonal elements of the jackknife covariance matrix for each bin.}
\label{fig:minimiser_accuracy_test}
\end{figure}

The black dots in Fig.~\ref{fig:minimiser_accuracy_test} represent the $w_{p}$ of the best-fitting HOD mock, with error bars derived from the jackknife covariance matrix. The red line shows the emulator prediction of $w_{p}$ for the true parameters used to construct the mock. As expected from our analysis, the emulator prediction differs slightly from the true $w_{p}$ due to inherent emulator errors, but this difference is less than 1~per~cent across all $r$ bins. The blue line corresponds to the best-fitting value determined by the minimization algorithm.  Because of the emulator errors, this prediction achieves a better fit (lower $\chi^{2}$) to the black dots than the true values. Nonetheless, within the error margins, both models are indistinguishable, as their differences are smaller than the uncertainties derived from the covariance matrix, as illustrated in the bottom panel of the figure. The values found by the minimizer are very similar to the true parameters. It is also reassuring that, out of the ten independent Powell minimizations we performed (see Section~\ref{HOD_bestfit}), approximately half converge to viable solutions that are very close to the true values in parameter space and have acceptable $\chi^{2}$ values. With this in mind, we believe our minimization methodology works as intended.

\bsp	
\label{lastpage}
\end{document}